\documentclass[article]{elsarticle}
\usepackage[margin=2.5cm]{geometry}
\usepackage{hyperref}
\hypersetup{
	colorlinks=true,
	linkcolor=blue,
	filecolor=black,      
	urlcolor=blue,
}
\pdfoutput=1 

\usepackage[utf8]{inputenc}
\journal{.}
\usepackage{amsmath}
\usepackage[thinlines]{easytable} 
\usepackage{wrapfig}
\numberwithin{equation}{section}

\def\non{\nonumber\\}

\def\e{\,{\rm e}}

\def\veps#1{\varepsilon_{#1}}

\def\kk#1#2{k_{#1}\cdot k_{#2}}

\def\half{{1\over 2}}

\def\Z{{\mathchoice {\hbox{$\sf\textstyle Z\kern-0.4em Z$}}
{\hbox{$\sf\textstyle Z\kern-0.4em Z$}}
{\hbox{$\sf\scriptstyle Z\kern-0.3em Z$}}
{\hbox{$\sf\scriptscriptstyle Z\kern-0.2em Z$}}}}

        \def\slash#1{#1\!\!\!\raise.15ex\hbox {/}}
\newcommand{\slD}{\,\raise.15ex\hbox{$/$}\kern-.27em\hbox{$\!\!\!D$}}
\newcommand{\slpartial}{\raise.15ex\hbox{$/$}\kern-.57em\hbox{$\partial$}}

\def\epsk#1#2{\varepsilon_{#1}\cdot k_{#2}}

\def\Gd{\dot{G}}

\def\no{\noindent}

\def\be{\begin{equation}}
\def\ee{\end{equation}\noindent}
\def\bear{\begin{eqnarray}}
\def\ear{\end{eqnarray}\noindent}
\def\bec{\blue\begin{equation}}
\def\eec{\end{equation}\black\noindent}
\def\bearc{\blue\begin{eqnarray}}
\def\earc{\end{eqnarray}\black\noindent}
\def\benn{\begin{enumerate}}
\def\enn{\end{enumerate}}
\def\ee{&=&}

\def\e{\,{\rm e}}

\def\b0{{\bf 0}}

\def\4piTD{{(4\pi T)}^{-{D\over 2}}}
\def\4piT4{{(4\pi T)}^{-2}}

\newcommand{\x}{{\hat{k}_{12}}}

\newcommand{\Gcc}{{\hat{\Gamma}_{{\rm scal}(34)}}}
\newcommand{\Gss}{{\hat{\Gamma}_{{\rm spin}(34)}}}

\newcommand{\Qcc}{Q_{{\rm scal}(34)}}
\newcommand{\Qss}{Q_{{\rm spin}(34)}}

\newcommand{\s}{{\rm scal}}
\newcommand{\sq}{{\rm spin}}

\newcommand{\glc}[1]{{\gamma_{{\rm scal}}^{#1}}}
\newcommand{\gls}[1]{{\gamma_{{\rm spin}}^{#1}}}

\usepackage[dvipsnames]{xcolor}

\begin{document}

\begin{frontmatter}
\title{The QED four-photon amplitudes off-shell: part 2}

\author[a]{Naser Ahmadiniaz}
\ead{n.ahmadiniaz@hzdr.de}
\author[b]{Cristhiam Lopez-Arcos}
\ead{cmlopeza@unal.edu.co}
\author[a]{Misha A. Lopez-Lopez}
\ead{m.lopez-lopez@hzdr.de}
\author[c,d]{Christian Schubert\corref{correspondingauthor}}
\cortext[correspondingauthor]{Corresponding author}
\ead{Christian.Schubert@eli-beams.eu}

\address[a]{Helmholtz-Zentrum Dresden-Rossendorf, Bautzner Landstra\ss e 400, 01328 Dresden, Germany}
\address[b]{Escuela de Matem\'{a}ticas, Universidad Nacional de Colombia Sede Medell\'{i}n, Carrera 65 $\#$ 59A--110, Medell\'{i}n, Colombia}
\address[c]{ELI Beamlines Centre, Institute of Physics, Czech Academy of Sciences, Za Radnicki 835, 25241 Dolni B\v{r}e\v{zany}, Czech Republic}
\address[d]{After May 15, 2023:
Instituto de F\'isica y Matem\'aticas,
Universidad Michoacana de San Nicol\'as de Hidalgo,
Edificio C-3, Apdo. Postal 2-82
C.P. 58040, Morelia, Michoac\'an, M\'exico}

\begin{abstract}
This is the second one of a series of four papers devoted to a first calculation of the scalar and spinor QED four-photon amplitudes completely off-shell. 
We use the worldline formalism which provides a gauge-invariant decomposition for these amplitudes as well as compact integral representations.
It also makes it straightforward to integrate out any given photon leg in the low-energy limit, and in the present sequel we do this with two of the four photons.
For the special case where the two unrestricted photon momenta are equal and opposite the information on these amplitudes is also contained in the constant-field vacuum polarisation tensors,
which provides a check on our results.  
Although these amplitudes are finite, for possible use as higher-loop building blocks we evaluate all integrals in dimensional regularisation.  
As an example, we use them to construct the two-loop vacuum polarisation tensors in the low-energy approximation, 
rederive from those the two-loop $\beta$-function coefficients and analyse their anatomy with respect to the gauge-invariant decomposition. 
As an application to an external-field problem, we provide a streamlined calculation of the Delbr\"uck scattering amplitudes in the low-energy limit. 
All calculations are done in parallel for scalar and spinor QED.
\end{abstract}

\end{frontmatter}


\tableofcontents


\section{Introduction}\label{sec:intro}

\begin{figure}[h]
	\centering 
	\includegraphics[width=.27\textwidth]{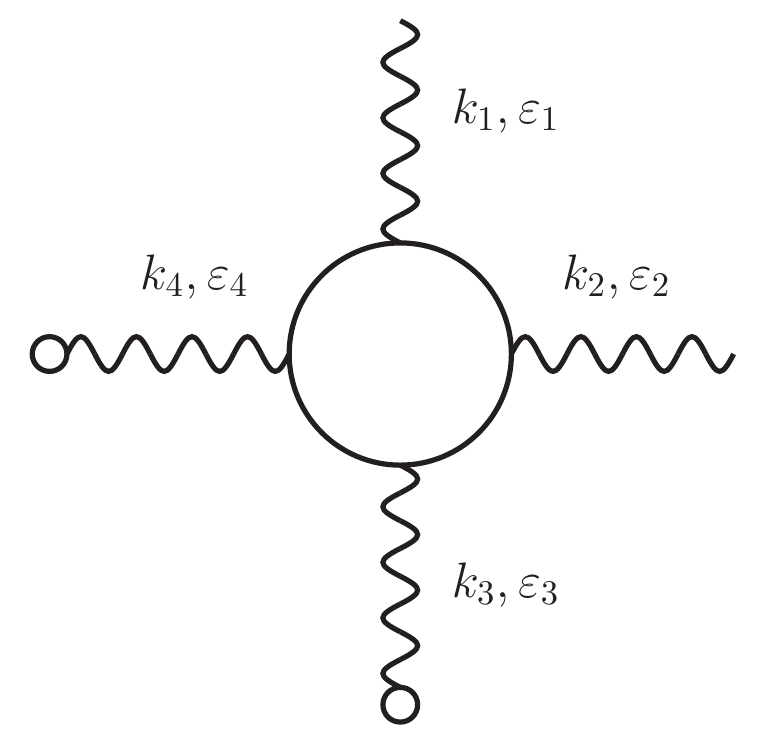}
	\caption{\label{43low}Four-photon box diagram with two low-energy legs $k_4$ and $k_3$ indicated by empty bullets at their ends. A sum over permutations
	is understood, as well as the inclusion of seagull diagrams in the scalar QED case. 
	 }
	\end{figure}

The present paper is the second one in a series of four devoted to a first calculation of the scalar and spinor QED 
one-loop four-photon amplitudes fully off-shell, using the worldline representation of these amplitudes. 
In part I \cite{Ahmadiniaz:2020jgo} we already derived these representations, discussed their properties, 
and calculated them for the simple limiting case where all four photons are taken in the low-energy (or ``Euler-Heisenberg'' limit). 
In the present sequel, we calculate them explicitly for the case where only photons 3 and 4 are taken in that limit, but 1 and 2 
have arbitrary off-shell momenta, see Fig. \ref{43low}. 

The results are given in compact form in terms of the hypergeometric function $_2F_1$ for the
dimensionally continued case, and in trigonometric form for $D=4$. 

Although to our knowledge this special case of the four-photon amplitudes has not been considered by other authors,
for $k_1=-k_2, k_3=-k_4$ it is straightforward to extract them from the well-studied photon polarization tensors in a constant field 
\cite{Batalin:1971au,Baier:1974qq,Baier:1974hn,Tsai:1975iz,Urrutia:1977fw,Schubert:2000yt,Dittrich2000zu,Karbstein:2015cpa}, 
and we use this fact for a check on our calculations. 

\begin{figure}[h]
		\centering 
		\includegraphics[width=.4\textwidth]{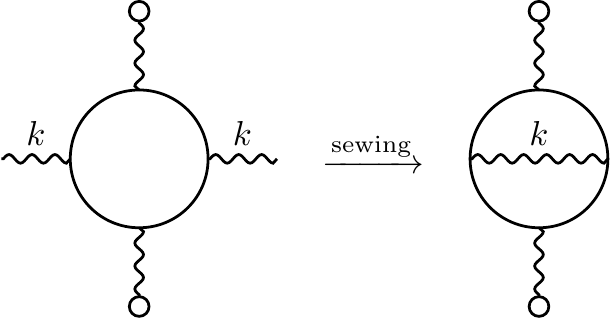}
		\caption{\label{beta} Construction of the two-loop photon propagator from the four-photon amplitude by sewing.}
	\end{figure}

Off-shell legs can be used for creating internal propagators by sewing, or for connecting them to external fields. 
As an example for the former, we use our results to construct, by sewing of the legs $1$ and $2$ that carry full momentum,
the two-loop scalar and spinor QED vacuum polarization tensors in the low-energy limit, see Fig. \ref{beta}. 

\begin{figure}[htbp]
\begin{center}
\includegraphics[scale=1.2]{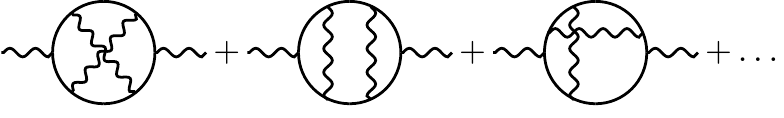}
\vspace{-.1cm}\caption{The three-loop quenched photon propagator.}
\label{fig-3loopbetadiag}
\end{center}
\end{figure}

In this limit the vacuum polarization tensors just reduce to the induced Maxwell terms, from which we can rederive the two-loop $\beta$-function coefficients. 
Apart from providing another check, this also improves on previous work \cite{Schmidt:1994aq} where
the worldline formalism had already been applied to the calculation of these coefficients. 
Although these coefficients were already known for decades \cite{joslut,biabirubetafunction}, that work was motivated by the fact that the formalism allows one
to obtain parameter-integral representations that unify all the Feynman diagrams contributing to the quenched 
(single fermion loop) photon propagator at any loop order, see., e.g., Fig. \ref{fig-3loopbetadiag} for the three-loop case.

\vspace{10pt}

\noindent
The interesting feature of the worldline formalism is that the sewing results in parameter integrals that
represent not some particular Feynman diagram, but the whole set of Feynman diagrams shown in Fig. \ref{fig-3loopbetadiag}. 
And it is precisely this type of sums of diagrams that are known for particularly extensive cancellations between diagrams.
In 1967 Johnson, Willey and Baker showed \cite{Johnson:1967pk} that the quenched vacuum polarisation has,
at any loop order, only a single overall UV divergence. And the coefficient of this pole, which is essentially the $\beta$-function,
turned out to be rational up to the four-loop level (see \cite{Broadhurst:1995dq} and refs. therein). 
The role of gauge invariance in these cancellations is not transparent, and remains an object of active investigation
even at the two-loop level \cite{grauel-thesis}. 
The worldline formalism was applied to the recalculation of the two-loop scalar and spinor QED $\beta$-functions by M.G. Schmidt and one of the authors 
 in \cite{Schmidt:1994aq} (see also \cite{Schubert2001-73}) to achieve significant cancellations already at the integrand level.
However, those calculations were done using a different approach based on two-loop worldline Green's functions
\cite{Schmidt:1994zj}, which provides a shortcut but obscures the role of gauge invariance.
Only after that work a refinement of the usual homogeneising Bern-Kosower integration-by-parts procedure was
 found that led to a 
decomposition of the four-photon amplitudes into sixteen individually gauge invariant contributions (see \cite{Schubert:1997ph,Ahmadiniaz:2012ie} and part I).
This gives us a chance to improve on \cite{Schmidt:1994aq} by asking the following two questions: first, can we identify
parts of that decomposition that drop out in the construction of the two-loop $\beta$ functions? 
Second, are there cancellations in their computation beyond what is expected from gauge invariance,
that is, not only inside each gauge invariant structure, but also between them?
Thus we will study here the distribution of the $\beta$-function coefficients, as well as the cancellation of the double pole in the $\frac{1}{\epsilon}$ expansion,
 in the gauge-invariant decomposition derived in part I. 

As an application to an external-field problem, 
among the many processes related to the four-photon amplitudes 
we have chosen here to reanalyse low-energy Delbr\"uck scattering,
the deflection of photons in the Coulomb field of nuclei due to the vacuum polarization 
(partially motivated also by the fact that the worldline formalism has already been applied extensively to
calculations involving constant 
\cite{Shaisultanov1996-354,Reuter1997-313,Schubert:2000yt,Schubert2001-73,Ahmadiniaz:2019nhk,Ahmadiniaz:2017rrk}, 
plane-wave \cite{Ilderton:2016qpj,Edwards:2021vhg}
and combinations of the two types of fields \cite{Schubert:2023gsl}, but not as yet to Coulomb fields). 
Delbr\"uck introduced this scattering in 1933 in order to explain the discrepancies Meitner and K\"osters had found in their experiment for the Compton scattering on heavy atoms \cite{meitner1933streuung}. 
Later Bethe and Rohrlich computed the angular distribution for small angles and the total cross section for Delbr\"uck scattering \cite{bethe1952small}. In 1973 DESY reported a first observation of this scattering in 
the high-energy, small-angle limit \cite{jarlskog1973measurement} which was in agreement with the prediction made in a series of papers by {\it Cheng} and {\it Wu} \cite{cheng1969high1,cheng1969high2,cheng1969high3,
cheng1969high4} and later confirmed in  \cite{mil1983quasiclassical,mil1983coherent}. 
In 1975 the G\"ottingen group performed another experiment which was the first one where exact predictions based on Feynman diagrams were confirmed with high precision after considering other background phenomena like atomic and nuclear Rayleigh scattering \cite{SCHUMACHER1975134}. The most accurate high energy  experiment done so far on Delbr\"uck scattering is the one carried out at BINP \cite{MILSTEIN1994183,SCHUMACHER1999101}. Here we will use our results for a short and efficient calculation of the Delbr\"uck scattering cross section in the low-energy limit. 

This paper is organized as follows. In Section \ref{WL_rep} we shortly review the worldline 
representation of the scalar and spinor QED  four-photon amplitudes, and discuss 
our general computational strategy in this paper. In particular, we provide all the integral formulas
required to integrate out low-energy photons without having to split these multi-photon integrals into ordered sectors. 
The central Section is \ref{2-low}, where we list our explicit final results for the amplitudes, 
in $D$ as well as in four dimensions, and for both scalar and spinor loops. 
Section \ref{check} contains the above-mentioned comparison with the amplitudes extracted from the vacuum polarization in a constant field.
In Section \ref{section-beta} 
we recalculate the two-loop $\beta$-functions, while Section \ref{delbruecksection} 
is devoted to low-energy Delbr\"uck scattering. 
Finally, Section \ref{summary} gives a summary and outline of future work. Supplementary information and formulas are provided in the appendices.

\section{Worldline representation of the four-photon amplitudes}
\label{WL_rep}
In part I, we presented in detail the derivation and structure of the worldline representation of the $N$-photon amplitudes. 
Due to the inherent freedom in the integration-by-parts procedure, this representation comes in various slightly different versions.
For easy reference, let us summarize here the representation of the four-photon amplitudes that we will actually use in the present sequel
\footnote{Since henceforth we are concerned exclusively with the four-photon case we will now omit the subscript $N$ on the $Q$'s.}:
%
%
\begin{equation}\label{expand4point-Q}
	\Gamma_{\rm scal}(k_1,\varepsilon_{1};\cdots;k_4,\varepsilon_4)
	=\frac{(-ie)^{4}}{(4\pi)^{\frac{D}{2}}}\int_{0}^{\infty} \frac{dT}{T}\, T^{4-\frac{D}{2}} \e^{-m^2 T}\, \int_0^1\prod_{i=1}^4du_i\, Q_{\s}\,\e^{(\cdot)}\,.
\end{equation}
Here we have already done the usual rescaling $\tau_i = Tu_i$ such that the exponential part is
\begin{equation}\label{i-q4}
	\e^{(\cdot)} \equiv \e^{T\sum_{i<j=1}^4G_{ij}k_i\cdot k_j}\,,	
\end{equation}
the bosonic Green's functions are
\begin{equation}
	G_{ij}\equiv G(u_i,u_j) =\vert u_i-u_j\vert-(u_i-u_j)^2 \,,
\end{equation}
and the polynomial $Q_{\s}$ is given by
\footnote{When comparing with \cite{Schubert2001-73} note that there a different basis was used for the four-cycle component $Q^4$. The two bases are related by cyclicity and inversion.}
\bear
Q_{\s}&=&Q^4_{\s}+Q^3_{\s}+Q^2_{\s}+Q^{22}_{\s}\,,\non
Q^4_{\s}&=&\Gd(1234)+\Gd(2314)+\Gd(3124)\,,\non
Q^3_{\s}&=&\Gd(123)T(4)+\Gd(234)T(1)+\Gd(341)T(2)+\Gd(412)T(3)\,,\non
Q^2_{\s}&=&\Gd(12)T_{sh}(34)+\Gd(13)T_{sh}(24)+\Gd(14)T_{sh}(23) +\Gd(23)T_{sh}(14)+\Gd(24)T_{sh}(13)+\Gd(34)T_{sh}(12)\,,\non
Q^{22}_{\s}&=&\Gd(12)\Gd(34)+\Gd(13)\Gd(24)+\Gd(14)\Gd(23)\,.\non
\label{q4-scal}
\ear
The extraordinary compactness of this representation is made possible through the introduction of the 
``Lorentz-cycle'' $Z_n(i_1i_2\ldots i_n)$ as 
\begin{eqnarray}
\begin{split}
Z_2(ij)&\equiv 
\half {\rm tr}\bigl(f_if_j\bigr) = \varepsilon_i\cdot k_j\varepsilon_j\cdot k_i - \varepsilon_i\cdot\varepsilon_jk_i\cdot k_j \, ,
\\
Z_n(i_1i_2\ldots i_n)&\equiv {\rm tr}
\Bigl(
\prod_{j=1}^n
f_{i_j}\Bigr)\, , 
\quad (n\geq 3) \, ,
\label{defZn}
\end{split}
\end{eqnarray}\no
where $f_i^{\mu\nu}=k_i^\mu \veps i^\nu-k_i^\nu\veps i^\mu$ is the field strength tensor of photon $i$,
and the ``bicycle'' 
\bear
\Gd(i_1i_2\cdots i_n)&\equiv& \Gd_{i_1i_2}\Gd_{i_2i_3}\cdots\Gd_{i_ni_1}Z_n(i_1\cdots i_n)\, .
\ear
It is the ``tails'' that exist in various versions. 
For the present computation, we use the one-photon tail $T(i)$ of the original $Q$-representation and the ``short tail'' $T_{sh}(ij)$, introduced in part I, as the two-photon tail:
\bear
T(i)&\equiv&\sum_{r\neq i}\Gd_{ir}\epsk ir\,,\\
T_{sh}(ij) &\equiv& \sum_{r,s\ne i,j}
\Gd_{ri} \Gd_{js}~ \frac{k_r \cdot f_i\cdot f_j\cdot k_s}{k_i\cdot k_j}\, .
\label{two-tails}
\ear
The spinor-loop result is obtained by employing the \emph{Bern-Kosower replacement rule}, i.e. replacing simultaneously 
every closed (full) cycle as $\Gd_{i_1i_2}\Gd_{i_2i_3}\cdots \Gd_{i_ni_1}$ appearing in the integrand of the scalar-loop with 
\bear
\Gd_{i_1i_2}\Gd_{i_2i_3}\cdots \Gd_{i_ni_1}-G_{Fi_1i_2}G_{Fi_2i_3}\cdots G_{Fi_ni_1}
\label{replacement} 
\ear
where $G_{Fij}={\rm sgn}(u_i-u_j)$ is the fermionic Green function
(here it is understood that $\dot G_{ij}=- \dot G_{ji}$ may have to be used to achieve the cycle form).  
We write the spinor-loop amplitude as
\bear
\Gamma_{\rm spin}(k_1,\varepsilon_{1};\cdots;k_4,\varepsilon_4)
 &=&-2\frac{(-ie)^{4}}{(4\pi)^{\frac{D}{2}}}\int_{0}^{\infty} \frac{dT}{T}\, T^{4-\frac{D}{2}} \e^{-m^2 T}\, \int_0^1\prod_{i=1}^4du_i\, Q_{\sq} \,\e^{(\cdot)}\,.
\label{expand4point-Q-spin}
\ear 
Thus, apart from a global factor of $-2$, the only difference to the scalar QED formula \eqref{expand4point-Q} is the replacement of $Q_{\rm scal}$ by $Q_{\rm spin }$ according to
the rule (\ref{replacement}). Let us also emphasize once more that equations (\ref{expand4point-Q}), (\ref{expand4point-Q-spin}) are valid off-shell, and that the right-hand sides
are manifestly finite term-by-term. The well-known spurious UV-divergences of the four-photon diagrams that usually cancel only in the sum of diagrams would show up here as
logarithmic divergences of the $T$-integration at $T=0$, but have been eliminated already at the beginning by the IBP procedure that led from the P-representation to the
Q-representation, see part I.

To avoid carrying common prefactors we define 
\begin{equation}
	\hat{\Gamma}_{\left\{\s\atop\sq\right\}}
	\equiv \int_{0}^{\infty} \frac{dT}{T}\, T^{4-\frac{D}{2}} \e^{-m^2 T}\, \int_0^1\prod_{i=1}^4du_i\, Q_{\left\{\s\atop\sq\right\}} \,\e^{(\cdot)}\,.
	\label{defGammahat}
\end{equation}

  \subsection{Case distinction for the low energy limit of photons $3$ and $4$}
 As has been mentioned above, in the present part II the photons number 3 and 4 are taken in low-energy limit which means that in the expressions $Q_{\rm scal/spin} \,e^{(\cdot)}$ we only consider those contributions which are linear in both $k_3$ and $k_4$ (for more details about the low-energy limit see part I). For the sake of clarity, in this subsection we list the different cases that appear in our calculations:
  \begin{itemize}
  \item {\bf Case 1}: Terms in $Q_{\rm scal}$ which are already linear in both $k_3$ and $k_4$ 
  do not need any further factors of $k_{3,4}$ from the exponential, so that we can simply replace $\e^{(\cdot)}\rightarrow \e^{(\cdot)\vert_{k_3,k_4\rightarrow 0}}=e^{TG_{12}\kk12}$. 
  This includes all the ``pure cycle'' terms, defined by the absence of tails. For instance, the four-cycle $Q_\s^4$
  \begin{equation}
  	\dot{G}(1234)\,\e^{(\cdot)}\rightarrow \dot{G}(1234)\e^{TG_{12}\kk12}.
  \end{equation}
  Or the two-two cycles  $Q_\s^{22}$, where it is possible to have both low-energy photons in the same cycle 
  \begin{equation}
  	\Gd(12) \Gd(34)\, \e^{(\cdot)}\rightarrow \Gd(12) \Gd(34)\,\e^{TG_{12}\kk12},
  \end{equation}
  or distributed among different cycles
  \bear
  \Gd(13) \Gd(24)\, \e^{(\cdot)}\rightarrow \Gd(13) \Gd(24)\,\e^{TG_{12}\kk12}.
  \ear
  
  \item {\bf Case 2}: Terms in $Q_\s^3$ which contain one of the low-energy momenta, say, $k_3$,
   in the three-cycle, but lack $k_4$. Those require an expansion of the exponential factor to linear order in $k_4$.
              For instance, the one-tail term $\Gd(123)T(4)\,\e^{(\cdot)}$ gives 
  \bear
  \Gd(123)T(4)\, \e^{(\cdot)}\rightarrow\Gd(123)\Big[\Gd_{41}\epsk41+\Gd_{42}\epsk42\Big]\Big[TG_{14}\kk14+TG_{24}\kk24\Big]\,\e^{TG_{12}\kk12}.
  \ear
  
  \item {\bf Case 3}: Terms in $Q_{\rm scal}^3$ which have two low-energy momenta in the three-cycle, for those we neglect terms in the one-tail that are linear in any of the two low-energy momenta. For instance, the one-tail term $\Gd(234)T(1)\,\e^{(\cdot)}$ gives 
  \bear
  \Gd(234)T(1)\, \e^{(\cdot)}\rightarrow \Gd(234)\Gd_{12}\epsk12 \e^{TG_{12}\kk12}.
  \ear
   However, it turns out that the contributions of these particular terms vanish after integration over $u_3$ and $u_4$.
  
  \item {\bf Case 4}: Terms in $Q_\s^2$ which lack both $k_3$ and $k_4$. 
    Such terms occur only in the two-tail term $\Gd(12)T_{\rm sh}(34)\,\e^{(\cdot)}$. 
    Here the exponential factor must be expanded to linear order in both $k_3$ and $k_4$,
  \bear
  \Gd(12)T_{\rm sh}(34)\, \e^{(\cdot)}\rightarrow\Gd(12)T_{\rm sh}(34) \left( TG_{34}\kk34 + T^2\sum_{i,j=1}^2 G_{i4}\kk{i}4 G_{j3}\kk{j}3 \right) \,\e^{TG_{12}\kk12}.
  \ear
   
  \item {\bf Case 5}: Terms in $Q_\s^2$ which have only one of the
  low-energy momenta in the two-cycle. For instance, the two-tail term $\Gd(13)T_{\rm sh}(24)\,\e^{(\cdot)}$ gives
  \bear
  \Gd(13)T_{\rm sh}(24)\, \e^{(\cdot)}\rightarrow  \Gd(13) \Gd_{12}\Gd_{41}\frac{k_1\cdot f_2\cdot f_4\cdot k_1}{\kk24} \Big[TG_{14}\kk14+TG_{24}\kk24\Big]\,\e^{TG_{12}\kk12} 
  \ear
  where we took the terms linear in $k_4$ from the exponential.

  \item {\bf Case 6}: Terms in $Q_\s^2$ which are already quadratic in at least one of the low-energy momenta can be discarded.  
  For instance, in the two-tail term $\Gd(34)T_{\rm sh}(12)\,\e^{(\cdot)}$ we have terms like 
  \bear
  \Gd(34)\Gd_{31}\Gd_{24}\frac{k_3\cdot f_1\cdot f_2\cdot k_4}{\kk12}\,\e^{(\cdot)}\rightarrow 0 \, .
  \ear
  In particular, all terms in $\Gd(34)T_{\rm sh}(12)\e^{(\cdot)}$ can be neglected.
  
  \end{itemize}

Thus we see that, in the limit of two low-energy photons which is our object of interest in this paper, there are many terms that drop out of the amplitude at the integrand level, and others that drop out after integration.
The transition to spinor QED does not lead to any new considerations. 

\subsection{Low-energy limit for two of the photons: computational examples}\label{caltwo}

Next, let us explain our strategy for computing the four-photon amplitude with two low-energy photons, using some sample terms from both 
$Q_{\s}$ and $Q_{\sq}$.

{\bf Example 1}: Let us consider the following one-tail term:	
\begin{equation}\label{q43,123}
\begin{split}
Q_{\s}^3(123;4) =&\textbf{ }\dot{G}_{12} \dot{G}_{23} \dot{G}_{31}Z_3(123) \left( \dot{G}_{41} \varepsilon_4 \cdot k_1+\dot{G}_{42} \varepsilon_4 \cdot k_2 +  \dot{G}_{43} \varepsilon_4 \cdot k_3 \right).
\end{split}
\end{equation}
In order to take the low-energy limit for leg 4, we define
\begin{equation}\label{Q3(4)}
Q^{3}_{\s(4)} (123;4) \equiv \left. \int_0^1 du_4 ~Q_{\s}^3(123;4)~ \e^{(\cdot 4)} \right|_{{\rm lin}~ k_4}\,,
\end{equation}
where in the full exponent we single out those terms which have a $k_4$
\bear 
\label{e(.4)}
\e^{(\cdot 4)} \equiv \e^{T(G_{14}\kk14+G_{24} \kk24+G_{34}\kk34)}\,,
\ear
such that 
\bear
\e^{(\cdot)} = \e^{(\cdot 4)}\,\e^{T(G_{12}\kk12+G_{13}\kk13+ G_{23}\kk23)}.
\ear
Here, for convenience, we set the following convention for low-energy legs: on the left-hand side of expression (\ref{Q3(4)}) the subscript `$(4)$' indicates that the leg number 4 
with momentum $k_4$ is to be taken in the low energy limit and integrated out. In  \ref{app2} we give a list of all the occurring integrals (up to permutations). 
It allows us to perform the integral over $u_4$ without fixing an ordering for the remaining photon legs (for efficient techniques to derive such formulas see, e.g., \cite{Edwards:2021elz}).

Since in this example we do not have linear terms in $k_4$, we take them from the exponential part and we use Eq. (\ref{intGs}) to integrate
\begin{equation}
\begin{split}
Q_{\s(4)}^3(123;4) & = \frac{T}{3} Z_3(123)\, \dot{G}_{12} \dot{G}_{23} \dot{G}_{31}\left(\sum_{i=1}^3 \dot{G}_{1i}G_{i1}\, k_i \cdot k_4 \varepsilon_4 \cdot k_1\right.\\
& \quad +\sum_{i=1}^3 \left.\dot{G}_{2i}G_{i2} \, k_i \cdot k_4 \varepsilon_4 \cdot k_2 + \sum_{i=1}^3 \dot{G}_{3i}G_{i3} \, k_i \cdot k_4 \varepsilon_4 \cdot k_3\right) \,.
\end{split}
\end{equation}
Combining terms, we can write the result in a manifestly gauge invariant form
\begin{equation}
\begin{split}
Q_{\s(4)}^3(123;4) &= \frac{T}{3} Z_3(123) \dot{G}_{12} \dot{G}_{23} \dot{G}_{31}\Big(\dot{G}_{12}G_{21} \, k_2 \cdot f_4 \cdot k_1 \\
&\quad + \dot{G}_{23}G_{32}\, k_3 \cdot f_4 \cdot k_2 +  \dot{G}_{31}G_{13}\,k_1 \cdot f_4 \cdot k_3  \Big) \,.\\
\end{split}
\label{q43scalar}
\end{equation}
%

Next we repeat all this with photon number 3. We define
\begin{equation}
	\begin{split}
		\label{Q3(34)}
		\Qcc^{3} (123;4) &\equiv \left. \int_0^1 du_3 \int_0^1 du_4 ~Q_{\s}^3(123;4) ~ \e^{(\cdot 3)}\e^{(\cdot 4)} \right|_{{\rm lin}~ k_4, k_3}\\
		&= \left. \int_0^1 du_3  ~Q_{\s(4)}^3(123;4) ~ \e^{(\cdot 3)} \right|_{{\rm lin}~ k_3}\, ,
	\end{split}
\end{equation}
where
\begin{equation}
	 \label{.3}
	 \e^{(\cdot 3)} \equiv \e^{T(G_{13}\kk13+G_{23}\kk23)}\,.
\end{equation}
Similarly to the above, the subscript `$(34)$' in equation (\ref{Q3(34)}) indicates that legs number 3 and 4 are both projected on their low energy limits. Note that some terms in (\ref{q43scalar}) are quadratic in $k_3$ 
and can be neglected. We use (\ref{intGdots}) to perform the integral over $u_3$ and, with the aid of the identity $\dot G_{ij}^2 = 1-4G_{ij}$, write the result as
\begin{equation}
\Qcc^3(123;4) = -\frac{T}{9} Z_3(123) k_2 \cdot f_4 \cdot k_1 \left(G_{12} - 10 G_{12}^2 + 24 G_{12}^3\right).
\label{example1}
\end{equation}

Therefore at this stage the contribution of $Q_{\s}^3(123;4)$ to the four-photon amplitude with legs 3 and 4 at low energy is given by 
\begin{equation}\label{ynl1}
\hat{\Gamma}_{{\rm scal}(34)}^3(123;4) =\int_{0}^{\infty} \frac{dT}{T}\, T^{4-\frac{D}{2}}\e^{-m^2 T}\,\int_0^1 du_1 du_2\, \Qcc^3(123;4)\,\e^{TG_{12} k_1\cdot k_2}\,.
\end{equation}
This leads us to define
\begin{equation}\label{defYnl}
\begin{split}
Y_{nl} \equiv \int_0^{\infty} \frac{dT}{T}\, T^{n-D/2} ~\int_0^1 du_1\int_0^1 du_2 ~ G_{12}^l \e^{-T[m^2 - G_{12} k_1 \cdot k_2]} \,.
\end{split}
\end{equation}
The proper-time integral $T$ is elementary and, due to the unbroken translation invariance along the loop,
 one of the two parameters $u_1,u_2$ can be fixed at some arbitrary value, such as $u_2 = 0$ and $u_1 = u$. 
 To the remaining $u$ integral we apply a tensor reduction procedure, explained in  \ref{B.2}, to arrive at the following expression,
\begin{equation}\label{Ynl}
\begin{split}
Y_{nl} = \frac{\Gamma\left(n-\frac{D}{2}-l\right)}{m^{2n-D} } ~\frac{d^l}{d\hat{k}_{12}^l} ~ {}_2F_1 \left(1,n-l-\frac{D}{2}; \frac{3}{2};\frac{\hat{k}_{12}}{4} \right),
\end{split}
\end{equation}
where $\hat{k}_{12} \equiv \frac{k_1\cdot k_2}{m^2}$ and ${}_2F_1(a,b;c;x)$ is the  Gauss hypergeometric function. 
With these definitions, \eqref{example1} and (\ref{ynl1}) transform into
\begin{equation}
\hat{\Gamma}_{{\rm scal}(34)}^3(123;4) = -\frac{1}{9}Z_3(123)\, k_2 \cdot f_4 \cdot k_1 \left(Y_{51} - 10 Y_{52} + 24 Y_{53}\right) \, .
\label{example1scal}
\end{equation}

Proceeding to the spinor QED case, since for the term at hand the integration over $u_4$ does not interfere with the
application of the Bern-Kosower replacement rule \eqref{replacement}, the term corresponding to (\ref{q43scalar}) reads 
\bear
\begin{split}
Q_{\sq(4)}^3(123;4) &= \frac{T}{3}\Big(\dot{G}_{12} \dot{G}_{23} \dot{G}_{31}-G_{F12}G_{F23}G_{F31}\Big)Z_3(123)\\
&\times\Big(\dot{G}_{12}G_{21}k_2 \cdot f_4 \cdot k_1 + \dot{G}_{23}G_{32}k_3 \cdot f_4 \cdot k_2 +  \dot{G}_{31}G_{13}k_1 \cdot f_4 \cdot k_3  \Big) \, .
\label{example1spinor}
\end{split}
\ear
It would seem that we now have to extend our table of integrals in \ref{app2} to integrals involving both
$\dot G_{ij}$ and $G_{Fij}$, but we can avoid this by eliminating the latter in favor of the former. 
This can be done using the basic identities
\bear
G_{Fij}^2 =1, \quad G_{Fij}G_{Fjk}G_{Fki} = -\dot{G}_{ij}-\dot{G}_{jk}-\dot{G}_{ki} 
\label{idbasic}
\ear
which transforms \eqref{example1spinor} into
\bear
\begin{split}
Q_{\sq(4)}^3(123;4) 
&=\frac{T}{3}\Big(\dot{G}_{12} \dot{G}_{23} \dot{G}_{31}+\dot{G}_{12}+\dot{G}_{23}+\dot{G}_{31}\Big)Z_3(123)\\
&\times\Big(\dot{G}_{12}G_{21}k_2 \cdot f_4 \cdot k_1 + \dot{G}_{23}G_{32}k_3 \cdot f_4 \cdot k_2 +  \dot{G}_{31}G_{13}k_1 \cdot f_4 \cdot k_3  \Big) \, .
\label{q43spin}
\end{split}
\ear
Proceeding as in the scalar case, one finds that the spinor-QED equivalent of \eqref{example1scal} becomes
\bear
	\Gss^3(123;4) &=& \frac{2}{9} Z_3(123) k_2 \cdot f_4 \cdot k_1\left(Y_{51} -Y_{52} - 12Y_{53}\right) \, ,
\ear
as the reader can easily verify using formulas of \ref{app2}.

{\bf Example 2}: Let us consider another one-tail term,
\begin{equation}\label{qf1}
\begin{split}
Q_{\s}^3(234;1)  & = Z_3(234)\, \dot{G}_{23} \dot{G}_{34} \dot{G}_{42} \left( \dot{G}_{12} \varepsilon_1 \cdot k_2+\dot{G}_{13} \varepsilon_1 \cdot k_3 +  \dot{G}_{14} \varepsilon_1 \cdot k_4 \right) \,.
\end{split}
\end{equation}
Notice that it has a quadratic term in $k_4$ and two linear terms in $k_4$. Omitting the former, and integration the latter over $u_4$ using (\ref{intGdots}), we obtain
\begin{equation}
Q_{\s(4)}^3(234;1) = Z_3(234)\, \dot{G}_{23} \left( \frac{1}{6}-\frac{1}{2}\dot{G}_{32}^2 \right) \left( \dot{G}_{12} \varepsilon_1 \cdot k_2+\dot{G}_{13} \varepsilon_1 \cdot k_3 \right)\,.
\end{equation}
Thus we have a term quadratic in $k_3$ and a term linear in $k_3$, and for the latter the integral over $u_3$ turns out to vanish. Therefore we get nothing, 

\begin{equation}
\Gcc^3(234;1) = 0\,.
\end{equation}
We leave it to the reader to check that the additional terms generated by applying the Bern-Kosower replacement rule to the $(234)$ cycle vanish, too.
Therefore
\begin{equation}
\Gss^3(234;1) = 0\,.
\end{equation}

{\bf Example 3}: As our final example, we consider the following 4-cycle term in the spinor QED integrand, 
\begin{equation}
Q_{\sq}^4(1234) = Z_4(1234) \Big(\dot{G}_{12} \dot{G}_{23} \dot{G}_{34} \dot{G}_{41} -G_{F12} G_{F23} G_{F34} G_{F41} \Big) \,.
\end{equation}
To reexpress the spin contribution $G_{F12} G_{F23} G_{F34} G_{F41}$, we insert a factor $1=-G_{F13}G_{F31}$, 
and once more use the identity $G_{Fij}G_{Fjk}G_{Fki} = -\dot{G}_{ij}-\dot{G}_{jk}-\dot{G}_{ki}$. This leads to
\begin{equation}
\begin{split}
Q_{\sq(4)}^4(1234) &= Z_4(1234) \int_0^1 du_4 \Big[\dot{G}_{12} \dot{G}_{23} \dot{G}_{34} \dot{G}_{41} 
+ \left(\dot{G}_{12} + \dot{G}_{23} + \dot{G}_{31} \right)\left( \dot{G}_{13}  + \dot{G}_{34} + \dot{G}_{41} \right) \Big] \,.
\end{split}
\end{equation}
Upon integration we find, after some cancellations,
\begin{equation}
\begin{split}
Q_{\sq(4)}^4(1234) = Z_4(1234) \left[\dot{G}_{12} \dot{G}_{23}\left( \frac{1}{6}-\frac{1}{2}\dot{G}_{31}^2 \right) + \left(\dot{G}_{12} + \dot{G}_{23} + \dot{G}_{31} \right)\dot{G}_{13}\right]\,.
\end{split}
\end{equation}
Integrating over $u_3$ we obtain
\begin{equation}
Q_{\sq(34)}^4(1234) = - \frac{4}{3} Z_4(1234) \left(G_{12}+2G_{12}^2 \right) \,.
\end{equation}
Thus with the notation (\ref{defYnl}), the contribution to the amplitude of this term reads
\begin{equation}
\Gss^4(1234) = - \frac{4}{3} Z_4(1234) \left(Y_{41}+2Y_{42} \right) \,.
\end{equation}

\section{Results: low-energy limit for two of the photons}
	\label{2-low}
	
	In this section, we present our results for the complete four-photon amplitudes in both scalar and spinor QED, 
	off-shell but under the restriction that photons 3 and 4 are taken in the low-energy limit i. e., we consider only contributions that are linear in $k_3$ and $k_4$ (see part I). 
	These amplitudes represent the main result of the present article. In the worldline formalism, they appear naturally decomposed as
	\begin{equation}
	 \hat \Gamma_{\left\{\s\atop\sq\right\}(34)}      = 
	 \hat \Gamma_{\left\{\s\atop\sq\right\}(34)}^4 
	 +
	  \hat \Gamma_{\left\{\s\atop\sq\right\}(34)}^3
	  +
	   \hat \Gamma_{\left\{\s\atop\sq\right\}(34)}^2
	   +
	    \hat \Gamma_{\left\{\s\atop\sq\right\}(34)}^{22}
	    \label{decomp}
	\end{equation}
	with
	\begin{equation}\label{Intu1u2T}
	 \hat \Gamma_{\left\{\s\atop\sq\right\}(34)}^i
	= \int_0^1 du_1 \int_0^1 du_2 \int_0^{\infty} \frac{dT}{T}\, T^{4-\frac{D}{2}}\e^{-T(m^2 - G_{12} k_1 \cdot k_2)} 
	 Q_{\left\{\s\atop\sq\right\}(34)}^i
	~~,~~i=4,3,2,22\,.~~~
	\end{equation}
Recall that the subscript $(34)$ in $\hat\Gamma_{(34)}^i$ indicates that the integrals over $u_3$ and $u_4$ were performed taking the low-energy limit for the corresponding photons. 
The absolute normalizations of the amplitudes are given by
	\bear
	\Gamma_{\rm scal(34)}(k_1,\varepsilon_{1}; \cdots;k_4,\varepsilon_4)
	&=&\frac{e^{4}}{ (4\pi)^{\frac{D}{2}}} \hat{\Gamma}_{\rm scal(34)} \, , \\
	\Gamma_{\rm spin(34)}(k_1,\varepsilon_{1}; \cdots;k_4,\varepsilon_4)
	&=&-2\frac{e^{4}}{ (4\pi)^{\frac{D}{2}}} \hat{\Gamma}_{\rm spin(34)}\,.
	\label{expand4point-34}
	\ear

	We give them first in dimensionally regularized form,
	and then in $D=4$. Since the worldline representation of these amplitudes is manifestly finite,
	taking this limit is trivial and will not involve a cancellation of spurious $1/\epsilon$ poles as would be the case in a 
	Feynman diagram calculation (in our conventions $\epsilon = D-4$).

\subsection{Dimensionally regularized amplitudes}

The integrals appearing in (\ref{Intu1u2T}) are all of the form of (\ref{defYnl}) so that we choose to express them in terms of the functions $Y_{nl}$
that have been given in terms of ${}_2F_1$ in (\ref{Ynl}). We note that this implies some redundancy since, as shown in  \ref{B.2}, 
there exist recursion relations between the $Y_{nl}$ functions that would also allow us to rewrite all our results entirely in terms of the $Y_{n0}$. 
However, the resulting representation would be less compact than the one given in the following. 
	
\subsubsection{Scalar QED}
	\label{two-two-s}
For $Q_{\s}^4$
	\bear
	\Gcc^4(1234) &=& \frac{2}{3} Z_4(1234) \left(Y_{41} -4 Y_{42} \right)\,,\\
	\Gcc^4(2314) &=& \frac{1}{9} Z_4(2314) \left(Y_{40}-12 Y_{41} + 36 Y_{42} \right)\,, \\
	\Gcc^4(3124) &=& \frac{2}{3} Z_4(3124) \left(Y_{41}-4 Y_{42} \right)\,.
	\ear
For $Q_{\s}^3$
	\bear
	\Gcc^3(123;4) &=& -\frac{1}{9}Z_3(123) k_2 \cdot f_4 \cdot k_1 \left(Y_{51} - 10 Y_{52} + 24 Y_{53}\right)\,,\\
	\Gcc^3(234;1) &=& 0\,,\\
	\Gcc^3(341;2) &=& 0\,,\\
	\Gcc^3(412;3) &=&-\frac{1}{9} Z_3(412) k_2 \cdot f_3 \cdot k_1 \left(Y_{51} - 10 Y_{52} + 24 Y_{53}\right)\,.
	\ear
For $Q_{\s}^2$
	\begin{equation}
	\begin{split}
	\Gcc^2(12;34) &= -\frac{1}{18} Z_2(12)  \Bigg\{ \left[\frac{1}{5}\left(Y_{50}- 4Y_{51} \right) - 6 \left(Y_{52}- 4Y_{53} \right)\right] k_1 \cdot f_3 \cdot f_4 \cdot k_2 \\
	& \quad + \frac{1}{5}\left(Y_{50}- 4Y_{51} \right) k_1 \cdot f_3 \cdot f_4 \cdot k_1 +(1\leftrightarrow 2) \Bigg\}\\
	& \quad -\frac{1}{9} Z_2(12) \left(Y_{62}- 8Y_{63} + 16 Y_{64} \right)  k_1 \cdot f_3 \cdot k_2 k_1 \cdot f_4 \cdot k_2 \,,
	\end{split}
	\end{equation}
	\begin{equation}
	\Gcc^2(13;24) + \Gcc^2(23;14)  = -\frac{1}{9} Z_2(13) k_1 \cdot f_2 \cdot f_4 \cdot k_1 \,  \left(Y_{51}- 4Y_{52} \right) + (1\leftrightarrow 2)\,,
	\end{equation}
	\begin{equation}
	\Gcc^2(14;23) + \Gcc^2(24;13)= - \frac{1}{9} Z_2(14) k_1 \cdot f_2 \cdot f_3 \cdot k_1 \,  \left(Y_{51}- 4Y_{52} \right) + (1\leftrightarrow 2)\,,
	\end{equation}
	\begin{equation}
	\Gcc^2(34;12) = 0\,.
	\end{equation} 
And finally for $Q_{\s}^{22}$
	\bear
	\Gcc^{22}(12,34) &=& \frac{1}{3}Z_2(12) Z_2(34) \left(Y_{40}-4Y_{41} \right)\,,\\
	\Gcc^{22}(13,24) &=&\frac{1}{9}Z_2(13)Z_2(24) Y_{40}\,,\\
	\Gcc^{22}(14,23) &=&\frac{1}{9}Z_2(14)Z_2(23) Y_{40}\,.  
	\ear

\subsubsection{Spinor QED}\label{2lowspinor}

	For $Q_{\sq}^4$
	\bear
	\Gss^4(1234) &=& - \frac{4}{3} Z_4(1234) \left(Y_{41}+2Y_{42} \right) \,,\\
	\Gss^4(2314) &=& - \frac{4}{9} Z_4(2314) \left(2Y_{40}- 6Y_{41} - 9 Y_{42} \right)\,,\\
	\Gss^4(3124) &=& - \frac{4}{3} Z_4(3124) \left(Y_{41}+2Y_{42} \right) \,.
	\ear
	For $Q_{\sq}^3$
	\bear
	\Gss^3(123;4) &=& \frac{2}{9} Z_3(123) k_2 \cdot f_4 \cdot k_1\left(Y_{51} -Y_{52} - 12Y_{53}\right)\,,  \\
	\Gss^3(234;1) &=& 0\,,\\
	\Gss^3(341;2) &=& 0\,,\\
	\Gss^3(412;3) &=& \frac{2}{9} Z_3(412) k_2 \cdot f_3 \cdot k_1 \left(Y_{51} - Y_{52} - 12Y_{53}\right)\,.
	\ear
	For $Q_{\sq}^2$
	\begin{equation}
	\begin{split}
	\Gss^2(12;34) &= \frac{2}{9} Z_2(12) \, \Bigg\{\left[\frac{1}{5}Y_{51}k_1 \cdot f_3 \cdot f_4 \cdot k_1 + \left(\frac{1}{5}Y_{51} - 6 Y_{53}\right)k_1 \cdot f_3 \cdot f_4 \cdot k_2   + (1\leftrightarrow 2) \right]\\
	& \quad + 2\, \left(Y_{63}- 4Y_{64} \right) k_1 \cdot f_3 \cdot k_2 k_1 \cdot f_4 \cdot k_2 \Bigg\}\,,
	\end{split}
	\end{equation}
	\begin{equation}
	\Gss^2(13;24) + \Gss^2(23;14) = \frac{2}{9} Z_2(13) k_1 \cdot f_2 \cdot f_4 \cdot k_1 \,  \left(Y_{51}- 4Y_{52} \right) + (1\leftrightarrow 2) \,,
	\end{equation}
	\begin{equation}
	\Gss^2(14;23) + \Gss^2(24;13)= \frac{2}{9} Z_2(14) k_1 \cdot f_2 \cdot f_3 \cdot k_1 \,  \left(Y_{51}- 4Y_{52} \right) + (1\leftrightarrow 2) \,,
	\end{equation}
	\begin{equation}
	\Gss^2(34;12) = 0\,.
	\end{equation}
	And finally for $Q_{\sq}^{22}$
	\begin{equation}
	\Gss^{22}(12,34) = \frac{8}{3} Z_2(12) Z_2(34) Y_{41}\,,
	\end{equation}
	\begin{equation}
	\Gss^{22}(13,24) =\frac{4}{9} Z_2(13)Z_2(24)Y_{40}\,,
	\end{equation}
	\begin{equation}
	\Gss^{22}(14,23) =\frac{4}{9}Z_2(14)Z_2(23) Y_{40}\,.
	\end{equation}

\subsection{Four-dimensional amplitudes}

	In this subsection we specialize the previous results to four space-time dimensions ($D=4$), after which it becomes possible to write the amplitudes completely in terms of elementary and
	  trigonometric functions. In the interest of compactness, it will now be useful to	introduce the following dimensionless variables,
	\bear \label{4-dim}
	\x=\frac{k_1\cdot k_2}{m^2}\,,
	~~~~~
	p_0 \equiv \frac{{\rm arcsinh} \Bigl(\frac{\sqrt{-\x}}{2} \Bigr)}{ \sqrt{(4-\x)(-\x)}}\,.
	\ear
	%

\subsubsection{Scalar QED}

For $Q_{\s}^4$ 
	\begin{equation}
		\begin{split}
			\Gcc^4(1234) &= -\frac{12 + 8(\x-6)p_0}{3m^4 \x^2}  Z_4(1234)\,,\\
			\Gcc^4(2314) &= -\frac{2(\x^2 -30\x +108)-8p_0(5\x^2 - 48\x + 108)}{9m^4(\x-4) \x^2} Z_4(2314)\,,\\
			\Gcc^4(3124) &= -\frac{12 + 8(\x-6)p_0}{3m^4 \x^2}  Z_4(3124)\,. 
		\end{split}
	\end{equation}
	For $Q_{\s}^3$
	\begin{equation}
		\begin{split}
			\Gcc^3(123;4) &= -\frac{2(\x^2 -48\x +180)-16p_0(4\x^2 - 39\x + 90)}{9m^6(\x-4) \x^3} Z_3(123) k_2 \cdot f_4 \cdot k_1 \,,\\
			\Gcc^3(412;3) &= -\frac{2(\x^2 -48\x +180)-16p_0(4\x^2 - 39\x + 90)}{9m^6(\x-4) \x^3} Z_3(412) k_2 \cdot f_3 \cdot k_1\,.
		\end{split}
	\end{equation}
	For $Q_{\s}^2$
	\begin{equation}
		\begin{split}
			\Gcc^2(12;34) &= \frac{-1}{90} Z_2(12)\Bigg\{\Bigg[ \frac{2(-\x + 2)-16p_0}{m^6(\x-4) \x} \, k_1 \cdot f_3 \cdot f_4 \cdot k_1 \\
			& + P_1 \, k_1 \cdot f_3 \cdot f_4 \cdot k_2  + (1\leftrightarrow 2)\Bigg] + 10 P_2 ~ k_2 \cdot f_4 \cdot k_1 k_2 \cdot f_3 \cdot k_1 \Bigg\}\,,\\
			\Gcc^2(13;24) &=  -\frac{2(\x-6)-16p_0(\x-3)}{9m^6 (\x-4)\x^2} Z_2(13)  k_1 \cdot f_2 \cdot f_4 \cdot k_1\,,\\
			\Gcc^2(23;14) &= -\frac{2(\x-6)-16p_0(\x-3)}{9m^6 (\x-4)\x^2} Z_2(23)   k_2 \cdot f_1 \cdot f_4 \cdot k_2 \,,\\
			\Gcc^2(14;23) &= -\frac{2(\x-6)-16p_0(\x-3)}{9m^6 (\x-4)\x^2} Z_2(14)   k_1 \cdot f_2 \cdot f_3 \cdot k_1 \,,\\
			\Gcc^2(24;31) &= -\frac{2(\x-6)-16p_0(\x-3)}{9m^6 (\x-4)\x^2} Z_2(24)   k_2 \cdot f_1 \cdot f_3 \cdot k_2 \,.
		\end{split}
	\end{equation}
And finally, for $Q_{\s}^{22}$
	\begin{equation}
		\begin{split}
			\Gcc^{22}(12,34) &= -\frac{2-8p_0}{3m^4 \x} Z_2(12) Z_2(34)\,,\\
			\Gcc^{22}(13,24) &= -\frac{2+8p_0}{9m^4(\x-4)} Z_2(13) Z_2(24)\,,\\
			\Gcc^{22}(14,23) &= -\frac{2+8p_0}{9m^4(\x-4)} Z_2(14) Z_2(23)\,.
		\end{split}
	\end{equation}
Here we have introduced two more functions, $P_i= P_i(\x)$ with $i=1,2$, which are defined as
	\begin{equation}
		P_1 \equiv \frac{2(- \x^3 +2 \x^2 -210\x + 900 ) -32p_0(8 \x^2 -90\x +225)}{m^6(\x-4) \x^3}\,,
	\end{equation}
	\begin{equation}
		P_2 \equiv \frac{4(-\x^2 + 55\x - 210) + 48p_0(3\x^2 - 30\x + 70)}{m^8(\x-4) \x^4}\,.
	\end{equation}

\subsubsection{Spinor QED}

	For $Q_{\sq}^4$
	\begin{equation}
		\begin{split}
			\Gss^4(1234) &= \frac{16[3+p_0(\x^2 + 2\x - 12 )]}{3m^4(\x-4) \x^2}  Z_4(1234)\,,\\
			\Gss^4(2314) &=  \frac{4[4\x^2 -3\x -54-8p_0(\x^2 + 3\x - 27 )]}{9m^4(\x-4) \x^2} Z_4(2314)\,,\\
			\Gss^4(3124) &= \frac{16[3+p_0(\x^2 + 2\x - 12 )]}{3m^4(\x-4) \x^2} Z_4(3124)\,. 
		\end{split}
	\end{equation}
For $Q_{\sq}^3$
	\begin{equation}
		\begin{split}
			\Gss^3(123;4) &= \frac{4(\x^2 +15\x-90)+16p_0(\x^2 - 30\x +90 )}{9m^6(\x-4)\x^3} Z_3(123) k_2 \cdot f_4 \cdot k_1 \,,\\
			\Gss^3(412;3) &= \frac{4(\x^2 +15\x-90)+16p_0(\x^2 - 30\x +90 )}{9m^6(\x-4)\x^3} Z_3(412) k_2 \cdot f_3 \cdot k_1\,.
		\end{split}
	\end{equation}
	For $Q_{\sq}^2$
	\begin{equation}
		\begin{split}
			\Gss^2(12;34) &= \frac{1}{3} Z_2(12)\Bigg\{\Bigg[ \frac{4(\x+2) + 32p_0(\x-1)}{15m^6(4-\x)^2\x}\, k_1 \cdot f_3 \cdot f_4 \cdot k_1 \\  
			&+\frac{2}{15}\tilde{P}_1 \, k_1 \cdot f_3 \cdot f_4 \cdot k_2 + (1\leftrightarrow 2) \Bigg] + \frac{4}{3}\tilde{P}_2\, k_2 \cdot f_4 \cdot k_1 k_2 \cdot f_3 \cdot k_1 \Bigg\}\,,\\
			\Gss^2(13;24) &= \frac{4(\x-6)-32p_0(\x-3)}{9m^6 (\x-4)\x^2} Z_2(13) k_1 \cdot f_2 \cdot f_4 \cdot k_1\,,\\
			\Gss^2(23;14) &= \frac{4(\x-6)-32p_0(\x-3)}{9m^6 (\x-4)\x^2} Z_2(23) k_2 \cdot f_1 \cdot f_4 \cdot k_2 \,,\\
			\Gss^2(14;23) &= \frac{4(\x-6)-32p_0(\x-3)}{9m^6 (\x-4)\x^2} Z_2(14) k_1 \cdot f_2 \cdot f_3 \cdot k_1\,,\\
			\Gss^2(24;31) &= \frac{4(\x-6)-32p_0(\x-3)}{9m^6 (\x-4)\x^2} Z_2(24) k_2 \cdot f_1 \cdot f_3 \cdot k_2\,.
		\end{split}
	\end{equation}
	And finally, for $Q_{\sq}^{22}$
	\begin{equation}
		\begin{split}
			\Gss^{22}(12,34)&= -\frac{16[1+2p_0(\x-2)]}{3m^4 (\x-4)\x} Z_2(12) Z_2(34)\,,\\
			\Gss^{22}(13,24) &=-\frac{8(1+4p_0)}{9m^4(\x-4)}  Z_2(13)Z_2(24)\,,\\	
			\Gss^{22}(14,23) &=-\frac{8(1+4p_0)}{9m^4(\x-4)}  Z_2(14)Z_2(23)\,,
		\end{split}
	\end{equation}
Here, we have defined $\tilde{P}_i= \tilde{P}_i(\x)$ as
	\bear
	\tilde{P}_1 \equiv \frac{2(\x^3 +32 \x^2 -390\x + 900)+16p_0(\x^3 -46 \x^2 + 270\x - 450 )}{m^6(\x-4)^2 \x^3},
	\ear 
	\bear
	\tilde{P}_2 \equiv \frac{2(- 23 \x^2 + 200\x - 420)-24p_0(\x^3 -18 \x^2 + 90\x - 140 )}{m^8(\x-4)^2 \x^4}.
	\ear 

\section{Check: the photon propagator in a constant field}
\label{check}

\begin{figure}[h]
		\centering 
	 	\includegraphics[width=.7\textwidth]{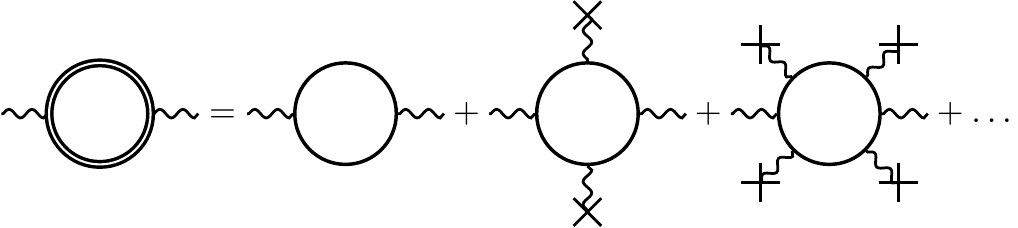}
		\caption{\label{vacuum} The first three terms in the diagrammatic expansion of the photon polarization tensor in powers of the constant background field. The
		second diagram on the right-hand side represents the four-photon amplitude with two low-energy photons and $k_1=-k_2$.}
	\end{figure}

The kinematical regime of the four-photon amplitudes that we are studying in this paper has, to the best of our
knowledge, not been treated in the literature before. However, for the special case where $k_1= - k_2$ (which 
then also implies $k_3 = -k_4$) they can be easily extracted from the photon polarization tensor in a constant background field,
a quantity that has been intensely studied in spinor 
\cite{Batalin:1971au,Baier:1974hn,Tsai:1975iz,Urrutia:1977fw,Schubert:2000yt,Dittrich2000zu,Karbstein:2015cpa}
and to a lesser degree in scalar QED \cite{Baier:1974qq,Schubert:2000yt}. 
The field can be treated non-perturbatively using the exact Dirac or Klein-Gordon propagator in the field, 
indicated by the double line on the left-hand side of Fig. \ref{vacuum} as is customary.

Expanding out the full non-perturbative polarization tensor in powers of the background field one obtains the series depicted on the right-hand side of
Fig. \ref{vacuum}, containing any even number of interactions with the field. Since a constant field cannot inject energy or momentum into the loop,
these interactions are mediated by zero-momentum photons. Thus the second diagram on the right-hand side of Fig. \ref{vacuum} corresponds
to our case of two arbitrary and two low-energy photons, the two low-energy legs corresponding to the two photons taken from the background field,
however under the restriction $k_1=-k_2=k$ imposed by the energy-momentum conservation in the constant field. 

We would now like to use this correspondence for a check on our above results. While such a comparison could be made using the results of
any of the references cited above, it will be convenient to compare with \cite{Schubert:2000yt}, where the worldline formalism was already used to
compute the constant-field photon polarization tensors for both scalar and spinor QED. Although that calculation is still quite different from our present
one (mainly because the integrating out of the low-energy photon legs there was effectively done already at the level of the construction of the
generalized worldline Green's functions, shown below in \eqref{defcalGB}, \eqref{defcalGF}), it is still close enough to our present one 
to make it possible to verify the equivalence at an early stage, without the need to perform all the integrals. It will be sufficient to work with 
the intermediate results of our above calculations after integrating out $u_3$ and $u_4$, which we collect in  \ref{app4}.
Using suitable integrations-by-part, we identify the resulting set of parameter integrals with the ones obtained by 
expanding the integral representations given in  \cite{Schubert:2000yt} for the scalar and spinor QED 
photon propagators in a constant background field to second order in the field strength tensor $F_{\mu\nu}$.
	
\subsection{Scalar QED}

	The above replacement for the momenta $(k_1=-k_2=k)$ actually makes some terms in $Q_{\s}$ vanish. The remaining terms in the scalar loop case can be compactly written as: 
	\begin{equation}\label{Q(34)in}
	\Qcc^4(1234) = \frac{2}{3}Z_4(1234) \left(G_{12}-4G_{12}^2\right)  ,
	\end{equation}
	\begin{equation}
	\Qcc^4(2314) = \frac{1}{9}Z_4(2314) \left(1- 12G_{12} + 36 G_{12}^2 \right) ,
	\end{equation}
	\begin{equation}
	\Qcc^4(3124) = \frac{2}{3} Z_4(3124) \left(G_{12} -4G_{12}^2\right)  , 
	\end{equation}
	\begin{eqnarray}
	\Qcc^3(123;4) = 0 , \\
	\Qcc^3(234;1) = 0 , \\
	\Qcc^3(314;2) = 0 , \\
	\Qcc^3(124;3) = 0 ,
	\end{eqnarray}
	\begin{equation}
	\Qcc^2(12;34) = - \frac{2T}{3}Z_2(12) k\cdot f_4 \cdot f_3 \cdot k \left(G_{12}^2 - 4G_{12}^3\right),
	\end{equation}
	\begin{equation}
	\Qcc^2(13;24) = - \frac{T}{9}Z_2(13) k\cdot f_4 \cdot f_2 \cdot k \left(G_{12} - 4G_{12}^2\right),
	\end{equation}
	\begin{equation}
	\Qcc^2(23;14) = - \frac{T}{9}Z_2(23) k\cdot f_4 \cdot f_1 \cdot k \left(G_{12} - 4G_{12}^2\right),
	\end{equation}
	\begin{equation}
	\Qcc^2(14;23) = -\frac{T}{9}Z_2(14)  k\cdot f_2 \cdot f_3 \cdot k \left(G_{12} - 4G_{12}^2\right) ,
	\end{equation}
	\begin{equation}
	\Qcc^2(24;13) = -\frac{T}{9}Z_2(24)  k\cdot f_1 \cdot f_3 \cdot k \left(G_{12} - 4G_{12}^2\right) ,
	\end{equation} 
	\begin{equation}
	\Qcc^2(34;12) = 0 ,
	\end{equation}
	\begin{equation}
	\Qcc^{22}(12,34) = \frac{1}{3} Z_2(12) Z_2(34) \left(1-4G_{12}\right) ,
	\end{equation}
	\begin{equation}
	\Qcc^{22}(13,24) =\frac{1}{9} Z_2(13)Z_2(24),
	\end{equation}
	\begin{equation}
	\Qcc^{22}(14,23) = \frac{1}{9} Z_2(14) Z_2(23). 
	\label{Q(34)fin}
	\end{equation}
	%
	Note that with the above replacement 			
	$f_{1}^{\mu\nu}=k^\mu\veps1^\nu-k^\nu \veps1^\mu$ and $f_{2}^{\mu\nu}=-k^\mu\veps2^\nu+k^\nu \veps2^\mu$.
	Rewriting the remaining integrals we have 
	\bear
	\Gamma_{\rm scal}(k,\veps1;-k,\veps2;k_3,\veps3;k_4,\veps4)=\frac{e^4}{(4\pi)^{\frac{D}{2}}}\int_0^\infty \frac{dT}{T}\, T^{4-\frac{D}{2}}\e^{-m^2T}\int_0^1du\,\Qcc \e^{-Tk^2G(u,0)}\,,\non
	\label{equalk-int}
	\ear
	where we have also set $u_2=0$ and $u_1=u$ so that now $G_{12}=G(u,0)=u(1-u)$. In order to be able to make the comparison at the integrand level is important to use IBP or equivalently the identity
	\begin{equation}\label{idGs}
	T \left(G_{12} - 4G_{12}^2 \right) k^2\,  \e^{-TG_{12} k^2}= \left(1-6 G_{12}\right) \e^{-TG_{12} k^2} -\partial_u \left(\Gd_{12} G_{12}\e^{-TG_{12} k^2}\right) 
	\end{equation}
	for many of the terms in $\Qcc^2$. Note that this identity can also be expressed in terms of the $Y_{nl}$ as 
	\bear
		\left(Y_{51} - 4 Y_{52}\right)k^2 =  Y_{40}- 6 Y_{41}
		\label{idYnl}
		\ear
		which will be useful in Section \ref{section-beta}.

	Turning our attention now to the photon polarization tensors in a constant background field, for scalar QED the worldline calculation of \cite{Schubert:2000yt} yielded the following presentation:
	\begin{equation}
		\Pi^{\mu\nu}_{\rm scal}=-\frac{e^2}{(4\pi)^{\frac{D}{2}}}\int_0^\infty dT \,T^{1-\frac{D}{2}}\e^{-m^2T}\,{\rm det}^{-\half}\left( \frac{\sin \mathcal{Z}}{\mathcal{Z}}\right) \int_0^1 du \, I^{\mu\nu}_{\rm scal}\e^{-Tk\cdot \Phi_{12}\cdot k}\,,
		\label{equal-k-constant-sc}
	\end{equation}
	where 
	\begin{equation}
		I_{\rm scal}^{\mu\nu}=\dot{\mathcal{G}}_{B12}^{\mu\nu}\,k\cdot\dot{\mathcal{G}}_{B12}\cdot k-\Big(\dot{\mathcal{G}}_{B11}-\dot{\mathcal{G}}_{B12}\Big)^{\mu\lambda}\Big(\dot{\mathcal{G}}_{B21}-\dot{\mathcal{G}}_{B22}\Big)^{\nu\kappa}k^{\lambda}k^\kappa \,,
		\label{equal-k-constant}
	\end{equation}
and ${\mathcal{G}}_{Bij}$ is the bosonic worldline Green's function in a constant field background \cite{Reuter1997-313},
	\begin{equation}
		\begin{split}
			\mathcal{G}_{Bij}&=\frac{T}{2\mathcal{Z}^2}\Big(\frac{\mathcal{Z}}{\sin \mathcal{Z}} \e^{-i\mathcal{Z}\dot{G}_{ij}}+i\mathcal{Z}\dot{G}_{ij}-1\Big)\,,\\
			\mathcal{G}_{Bii}&=\frac{T}{2\mathcal{Z}^2}\Big(\mathcal{Z}\cot \mathcal{Z}-1\Big)\,,
		\end{split}
		\label{defcalGB}
	\end{equation}
	with $\mathcal{Z}^{\mu\nu}=eTF^{\mu\nu}$. For the exponent of (\ref{equal-k-constant-sc}), we have 
	\begin{equation}
		-Tk\cdot \Phi_{ij}\cdot k=-k\cdot\left[\frac{T}{2\mathcal{Z}}\Big(\frac{\e^{-i\mathcal{Z}\dot{G}_{ij}}-\cos\mathcal{Z}}{\sin \mathcal{Z}}+i\dot{G}_{ij}\Big)\right]\cdot k \,.
	\end{equation}
	
	In order to compare with the results in the present work, we choose $F^{\mu\nu} = i(f_3^{\mu\nu} + f_4^{\mu\nu})$ and expand the whole polarization tensor
	to second order in $F^{\mu\nu}$ so as to obtain the first two diagrams on the right-hand side of Fig. \ref{vacuum}.  After some straightforward algebra, we can express the result as 
	\begin{equation}
		\begin{split}
			\varepsilon_1\cdot \Pi_{\rm scal}\cdot\veps2 &= - \frac{e^4}{(4\pi)^{D/2}} \int_0^{\infty} dT \, T^{3-D/2} \e^{-m^2 T} \int_0^1 du \, \e^{-TG_{12}k^2}\\
			&\hspace{-1cm}\times\bigg\{ (1-4G_{12}) \Big(\varepsilon_1 \cdot \varepsilon_2 k^2 - \varepsilon_1 \cdot k \varepsilon_2 \cdot k\Big) \Big[(eT)^{-2} -\frac{2T}{3}G_{12}^2 k\cdot f_3\cdot f_4 \cdot k + \frac{1}{6} tr(f_3f_4)  \Big] \\
			&\hspace{-0.8cm}+ \frac{2}{3} \Big(G_{12} -4G_{12}^2\Big) \Big[2\varepsilon_1 \cdot \varepsilon_2 k\cdot f_3\cdot f_4 \cdot k + k^2 \left(\varepsilon_1 \cdot f_3 \cdot f_4 \cdot \varepsilon_2 + \varepsilon_1 \cdot f_4 \cdot f_3\cdot \varepsilon_2 \right)\\
			&\hspace{-0.8cm}- \varepsilon_2 \cdot k\Big( \varepsilon_1 \cdot f_3 \cdot f_4 \cdot k + \varepsilon_1 \cdot f_4 \cdot f_3 \cdot k\Big) - \varepsilon_1 \cdot k\Big( \varepsilon_2 \cdot f_3 \cdot f_4 \cdot k + \varepsilon_2 \cdot f_4 \cdot f_3 \cdot k\Big)\Big] \\
			&\hspace{-0.8cm} - 4 G_{12}^2 \Big( \varepsilon_1 \cdot f_3 \cdot k  \varepsilon_2 \cdot f_4 \cdot k + \varepsilon_1 \cdot f_4 \cdot k  \varepsilon_2 \cdot f_3 \cdot k\Big) \bigg\}\,,
		\end{split}
	\end{equation}
	Note that the term proportional to $(eT)^{-2}$ corresponds to the two-point diagram or the vacuum polarization diagram in the absence of the background field. 
	The remaining terms give the sought after four-photon amplitude, and are found to be in complete agreement with (\ref{equalk-int}).

\subsection{Spinor QED}

	For the spinor QED case we find, in complete analogy,
	\begin{equation}
	\Qss^4(1234) = -\frac{4}{3} Z_4(1234) \left( G_{12}+ 2G_{12}^2\right) ,
	\end{equation}
	\begin{equation}
	\Qss^4(2314) =  -\frac{4}{9} Z_4(2314) \left(2- 6 G_{12}- 9G_{12}^2 \right) ,
	\end{equation}
	\begin{equation}
	\Qss^4(3124) = -\frac{4}{3} Z_4(3124) \left( G_{12}+ 2G_{12}^2\right) , 
	\end{equation}
	\begin{eqnarray}
	\Qss^3(123;4) = 0 , \\
	\Qss^3(234;1) = 0 , \\
	\Qss^3(314;2) = 0 , \\
	\Qss^3(124;3) = 0 , 
	\end{eqnarray}
	\begin{equation}
	\Qss^2(12;34) = \frac{8T}{3} Z_2(12) k\cdot f_4 \cdot f_3 \cdot k \, G_{12}^3 \,,
	\end{equation}
	\begin{equation}
	\Qss^2(13;24) =  \frac{2T}{9}Z_2(13) k\cdot f_4 \cdot f_2 \cdot k \left(G_{12} - 4G_{12}^2\right),
	\end{equation}
	\begin{equation}
	\Qss^2(23;14) = \frac{2T}{9}Z_2(23) k\cdot f_4 \cdot f_1 \cdot k \left(G_{12} - 4G_{12}^2\right),
	\end{equation}
	\begin{equation}
	\Qss^2(14;23) = \frac{2T}{9}Z_2(14) k\cdot f_2 \cdot f_3 \cdot k \left(G_{12} - 4G_{12}^2\right) ,
	\end{equation}
	\begin{equation}
	\Qss^2(24;13) = \frac{2T}{9}Z_2(24) k\cdot f_1 \cdot f_3 \cdot k \left(G_{12} - 4G_{12}^2\right) ,
	\end{equation} 
	\begin{equation}
	\Qss^2(34;12) = 0 ,
	\end{equation}
	\begin{equation}
	\Qss^{22}(12,34) = \frac{8}{3} Z_2(12) Z_2(34) G_{12}\,,
	\end{equation}
	\begin{equation}
	\Qss^{22}(13,24) =\frac{4}{9} Z_2(13)Z_2(24),
	\end{equation}
	\begin{equation}
	\Qss^{22}(14,23) = \frac{4}{9}  Z_2(14) Z_2(23),
	\end{equation}
	and the following integral representation 
	\begin{equation}
		\Gamma_{\rm spin}(k,\veps1;-k,\veps2;k_3,\veps3;k_4,\veps4)=-2\frac{e^4}{(4\pi)^{\frac{D}{2}}}\int_0^\infty \frac{dT}{T}\, T^{4-\frac{D}{2}}\e^{-m^2T}\int_0^1du\,\Qss \e^{-Tk^2G(u,0)}\,.\\
		\label{equalk-int-sp}
	\end{equation}
	As in the scalar case, to be able to compare at the integrand level it is important to make repeated use of (\ref{idGs}). 
	
	The spinor-loop result of the vacuum polarization tensor (from \cite{Schubert2001-73,Schubert:2000yt}) is
	\begin{equation}
		\Pi^{\mu\nu}_{\rm spin} = 2\frac{e^2}{(4\pi)^{\frac{D}{2}}}\int_0^\infty dTT^{1-\frac{D}{2}}\e^{-m^2T}\, {\rm det}^{-\half}\left(\frac{\tan \mathcal{Z}}{\mathcal{Z}}\right) \int_0^1 du\,I^{\mu\nu}_{\rm spin}\,\e^{-Tk\cdot \Phi_{12}\cdot k}\,,
		\label{equal-k-constant-sp}
	\end{equation} 
	where 
	\begin{equation}
		\begin{split}
			I_{\rm spin}^{\mu\nu}&=\dot{\mathcal{G}}_{B12}^{\mu\nu}\,k\cdot\dot{\mathcal{G}}_{B12}\cdot k-\mathcal{G}_{F12}^{\mu\nu}\,k\cdot\mathcal{G}_{F12}\cdot k\\
			&-\Big[\Big(\dot{\mathcal{G}}_{B11}-\mathcal{G}_{F11}- \dot{\mathcal{G}}_{B12}\Big)^{\mu\lambda} \Big(\dot{\mathcal{G}}_{B21}-\dot{\mathcal{G}}_{B22} +\mathcal{G}_{F22}\Big)^{\nu\kappa} +\mathcal{G}_{F12}^{\mu\lambda} \mathcal{G}_{F21}^{\nu\kappa}\Big]k^{\lambda} k^\kappa\,,
		\end{split}
	\end{equation}
	with the fermionic Green's function in a constant field and its coincidence limit \cite{Reuter1997-313}
	\begin{equation}
		\begin{split}
		\mathcal{G}_{Fij}&=G_{Fij}\frac{\e^{-i\mathcal{Z}\dot{G}_{ij}}}{\cos\mathcal{Z}}\,,\\
		\mathcal{G}_{Fii}&=-i\tan\mathcal{Z}\,.	
		\end{split}
		\label{defcalGF}
	\end{equation}
	After the expansion, choosing $\mathcal{Z}$ as in the scalar case, the spinor-loop gives 
	\begin{equation}
		\begin{split}
			\varepsilon_{1}\cdot\Pi_{\rm spin}\cdot\veps2& =  \frac{2e^4}{(4\pi)^{\frac{D}{2}}} \int_0^{\infty} dT \, T^{3-\frac{D}{2}} \e^{-m^2 T} \int_0^1 du \, \e^{-TG_{12}k^2}\\
			&\hspace{-1cm}\times\bigg\{ 4 G_{12} \Big(\varepsilon_1 \cdot \varepsilon_2 k^2 - \varepsilon_1 \cdot k \varepsilon_2 \cdot k\Big)\Big[-(eT)^{-2} +\frac{2T}{3} G_{12}^2 k\cdot f_3\cdot f_4 \cdot k+ \frac{1}{3} {\rm tr}(f_3f_4)  \Big]  \\
			&\hspace{-0.8cm}- \frac{4}{3} \Big(G_{12} + 2 G_{12}^2\Big) \Big[2\varepsilon_1 \cdot \varepsilon_2 k\cdot f_3\cdot f_4 \cdot k + k^2 (\varepsilon_1 \cdot f_3 \cdot f_4 \cdot \varepsilon_2 + \varepsilon_1 \cdot f_4 \cdot f_3 \cdot \varepsilon_2)\\
			&\hspace{-0.8cm}- \varepsilon_2 \cdot k\Big( \varepsilon_1 \cdot f_3 \cdot f_4 \cdot k + \varepsilon_1 \cdot f_4 \cdot f_3 \cdot k\Big) - \varepsilon_1 \cdot k\Big( \varepsilon_2 \cdot f_3 \cdot f_4 \cdot k + \varepsilon_2 \cdot f_4 \cdot f_3 \cdot k\Big) \Big]\\
			&\hspace{-0.8cm}-4  G_{12}^2 \Big( \varepsilon_1 \cdot f_3 \cdot k  \varepsilon_2 \cdot f_4 \cdot k + \varepsilon_1 \cdot f_4 \cdot k  \varepsilon_2 \cdot f_3 \cdot k\Big) \bigg\}\,,
		\end{split}
	\end{equation}
	which again after removing the term proportional to $(eT)^{-2}$ is in perfect agreement with (\ref{equal-k-constant-sp}). Despite of the restriction $k_1=-k_2$, this provides an excellent check on 
	many of the results presented in Section \ref{2-low} above.

\section{Gauge-invariant decomposition of the two-loop $\beta$-functions}
\label{section-beta}

	In this section, we use the results presented in Section \ref{2-low} to go to two loops by sewing together the unrestricted photons 1 and 2 (Fig. \ref{beta}).
	This yields the two-loop photon propagators in the low-energy limit, that is, the induced Maxwell terms, from whose $1/\epsilon$-poles we can (up to a contribution
	from mass renormalization) compute the two-loop $\beta$-function coefficients for scalar and spinor QED, recuperating the known results 	
	\cite{joslut,biabirubetafunction,Itzykson1980rh,Schmidt:1994aq,Schmidt1996-2150,Schubert2001-73}. 
	As explained in the introduction, besides providing another check this
	is motivated by the following two questions suggested by the highly structured worldline integrand \eqref{q4-scal}: first, are there terms that drop out in the sewing procedure? Second, does
	the fact that the sixteen terms in this decomposition are (differently from the usual decomposition into Feynman diagrams) individually gauge invariant, have a bearing on the issue
	of gauge cancellations? Specifically, will the vanishing of the double poles involve only cancellations inside each gauge-invariant structure, or between them?
	
	We use Feynman gauge in the sewing, so that it is done by the replacement 
	\begin{equation}
		\veps1^\mu\veps2^\nu\rightarrow \frac{\eta^{\mu\nu}}{k^2}\,,
	\end{equation}
	($k_1=k=-k_2$) with $k$ to be integrated over. The induced Maxwell term in our present set-up will appear as ${\rm tr}(f_3f_4)$.
The sewing will now change the argument of $Y_{nl}$ functions, such that in this section it is understood that $Y_{nl} = Y_{nl}(-k^2,m^2,D)$.
Going back to the results in $D$-dimensions in Section \ref{2-low} for the scalar/spinor-loop, we have the following expressions which will be used in both cases:
	\begin{equation}
		\begin{split}
			Z_4(1234)\vert_{\rm sewing}=Z_4(3124)\vert_{\rm sewing}&\Rightarrow\frac{2}{D}(D-1){\rm tr}(f_3f_4)\,,\\
			Z_4(2314)\vert_{\rm sewing}&\Rightarrow\frac{2}{D}{\rm tr}(f_3f_4)\,,\\
			Z_3(123)\vert_{\rm sewing}\, k\cdot f_4\cdot k &\Rightarrow0\,,\\
			Z_2(12)\vert_{\rm sewing}\, k\cdot f_3\cdot f_4\cdot k &\Rightarrow\frac{1}{D}(D-1)k^2{\rm tr}(f_3f_4)\,,\\
			Z_2(13)(k\cdot f_4\cdot f_2\cdot k)\vert_{\rm sewing}&\Rightarrow\frac{1}{D}{\rm tr}(f_3f_4)k^2\,,\\
			Z_2(12)Z_2(34)\vert_{\rm sewing}&\Rightarrow\frac{(D-1)}{2}{\rm tr}(f_3f_4)\,,\\
			Z_2(13)Z_2(24)\vert_{\rm sewing}=Z_2(14) Z_2(23)\vert_{\rm sewing}&\Rightarrow\frac{1}{D}{\rm tr}(f_3f_4)\,.\\
			\label{sewing-I}
		\end{split}
	\end{equation}
	Here we have used the Passarino-Veltman type identity 
	\begin{equation}
			\int d^D k\,(k\cdot f_3\cdot f_4\cdot k)\,  F\left(k^2\right)\rightarrow \int d^D k\,\frac{k^2}{D}{\rm tr}(f_3f_4) \, F\left(k^2\right)\,,
		\label{idPV}
	\end{equation}
	where $F$ is an arbitrary scalar function. 
	
	Note that the three-cycle $Z_3(123)$ drops out in the sewing, together with its three permutations. Therefore $Q^3_{\rm scal/spin}$ does not contribute here, and the total
	amplitudes can be written as (using a similar convention as in \eqref{defGammahat})
	\bear
	\begin{split}
	\Gamma^{(2)}_{\rm scal}\big(f_3,f_4\big)&= \frac{e^4}{2(4\pi)^{\frac{D}{2}}}\int\frac{d^Dk}{(2\pi)^D} \left(  \hat{\Gamma}^{(2)4}_{\rm scal} + \hat{\Gamma}^{(2)2}_{\rm scal} + \hat{\Gamma}^{(2)22}_{\rm scal} \right)\, , \\
\Gamma^{(2)}_{\rm spin}\big(f_3,f_4\big)&= -\frac{e^4}{(4\pi)^{\frac{D}{2}}}\int\frac{d^Dk}{(2\pi)^D} \left(  \hat{\Gamma}^{(2)4}_{\rm spin} + \hat{\Gamma}^{(2)2}_{\rm spin} + \hat{\Gamma}^{(2)22}_{\rm spin} \right)\,. 
	\label{defGammahat2loop}
	\end{split}
	\ear
	 (note that a symmetry factor of $\frac{1}{2}$ has to be included).

	\subsection{Scalar QED $\beta$-function}
\label{scalar-beta}

In this way, we find the following result for the Maxwell term of the two-loop vacuum polarization tensor in scalar QED,
	\begin{equation}
	\begin{split}
	\Gamma^{(2)}_{\rm scal}\big(f_3,f_4\big) &=\frac{e^4}{2(4\pi)^{\frac{D}{2}}} \int\frac{d^Dk}{(2\pi)^D} \Bigg\{\left(\frac{4}{9D} + \frac{D-1}{6} \right)Y_{40} - \frac{2}{3D}\big[(D-4)(D-1)+4\big]Y_{41}\\
	& -\frac{8}{3D}(4D-7)Y_{42}-\frac{4}{9D}(Y_{51}-4Y_{52})k^2 -\frac{2(D-1)}{3D}(Y_{52}-4Y_{53})k^2\Bigg\}{\rm tr}(f_3f_4) \,.
	\end{split}
	\end{equation}
Using the identity \eqref{idYnl} this can be somewhat simplified, 
	\begin{equation}
		\begin{split}
			\Gamma^{(2)}_{\rm scal}(f_3,f_4)&=\frac{e^4}{2(4\pi)^{\frac{D}{2}}}\int\frac{d^Dk}{(2\pi)^D}\bigg[\frac{D-1}{6}Y_{40}-\frac{2}{3D}(D-4)(D-1)Y_{41}\\
			&\hspace{1cm}-\frac{8}{3D}(4D-7)Y_{42}-\frac{2(D-1)}{3D}\, k^2(Y_{52}-4Y_{53})\bigg]{\rm tr}
			(f_3f_4)\,.
		\end{split}
	\end{equation}
Next, we rewrite this as 
	\bear
	\Gamma^{(2)}_{\rm scal}(f_3,f_4)\equiv\frac{e^4}{2(4\pi)^4} \gamma_{\s}(D,m^2){\rm tr} (f_3f_4)\,.
	\label{2loopbeta}
	\ear
	We then use (\ref{u-integrals}) to eliminate the $k$ - integration, which leads to
	\begin{equation}
		\begin{split}
			\hspace{-0.3cm}\gamma_{\s} (D,m^2) &= \frac{ (4 \pi)^{4}  \Gamma (4-D)}{4 (4 \pi)^{D}m^{2(4-D)}} \, \int_0^1 \,du \, \frac{1}{3D}
			\Big\{2 (D-1) D \Big[G(u,0)\Big]^{-\frac{D}{2}} \\
			& \quad -4(D-1) (3 D-8) \Big[G(u,0)\Big]^{1-\frac{D}{2}}  + 16 (D-7) (D-2) \Big[G(u,0)\Big]^{2-\frac{D}{2}}\Big\} .
		\end{split}
	\end{equation}
	 The remaining $u$-integral can be expressed in terms of the Euler Beta-function as
 \begin{equation}
 	\int_0^1du\Big[G(u,0)\Big]^x=B(x+1,x+1), \qquad  B(x,y) \equiv \frac{\Gamma(x)\Gamma(y)}{\Gamma(x+y)}\,,
 	\label{betfn}
 \end{equation}
	which leads to our final result for the coefficient of the induced Maxwell term,
		\begin{equation}
		\gamma_{\s} (D,m^2) = - \frac{(4\pi)^{4}\, \Gamma\left(4-D\right)}{(4\pi)^D \, m^{2(4-D)}}  \,  B\left(1-\frac{D}{2},1-\frac{D}{2}\right) \frac{(D-4) [(D-3) D+8]}{12 (D-5) (D-3) D} \, .
	\end{equation}
	 Note the global factor of $(D-4)$, which makes it clear that, as expected, the $1/\epsilon$-expansion has only a single pole,
	\bear
	\gamma_{\s}(\epsilon,m^2)&=&\frac{2}{\epsilon} +\mathcal{O} (\epsilon^0)\,.
	\label{beta-reg}
	\ear
After substituting (\ref{beta-reg}) into (\ref{2loopbeta}) one arrives at 
	\bear
	\Gamma^{(2)}_{\rm scal}(f_3,f_4)=\frac{e^4}{(4\pi)^4\epsilon} {\rm tr} (f_3f_4) +\mathcal{O} (\epsilon^0)\,,
	\ear
	which is in agreement with previous calculations of the two-loop dimensionally regularized effective action in Scalar QED \cite{Schmidt:1994aq,Fliegner:1997ra} 
	\footnote{Note that there is an overall sign difference between the above results and the one obtained in 
	\cite{Schubert2001-73} which comes from the fact that the two external photons in Fig. \ref{beta} have equal and opposite momenta ($k_3=-k_4$) which leads to an extra sign in ${\rm tr}(f_3f_4)$. } .

For the full two-loop renormalization of the photon propagator, one still needs to add an equal
 contribution $\Delta\Gamma^{(2)}_{\rm scal}(f_3,f_4)$ from mass renormalization \cite{Schmidt:1994aq},  
	\bear
	\Gamma^{(2)}_{\rm scal}(f_3,f_4)+\Delta\Gamma^{(2)}_{\rm scal}(f_3,f_4)\sim \frac{2e^4}{(4\pi)^4\epsilon}{\rm tr}(f_3f_4)\,.
	\ear
	This then gives the two-loop photon wave-function renormalization factor and leads to the standard result for the two-loop $\beta$-function coefficient 
	in Scalar QED \cite{biabirubetafunction,Schmidt:1994aq}
	\bear
	\beta^{(2)}_{\rm scal}(\alpha)=\frac{\alpha^3}{2\pi^2}~~~~,~~~~\alpha=\frac{e^2}{4\pi}\,.
	\ear

	More interesting is the distribution of the double and single - pole coefficients $a_{\rm scal}$ and $b_{\rm scal}$ in the decomposition \eqref{q4-scal}, shown in 
	 Table \ref{table:ta2}. 
	
	\begin{table}[h!]
		\centering
		\begin{tabular}{ |c|c|c||c|c|c||c|c|c| }
			\hline
			$\glc{}$ & $a_{\rm scal}$ & $b_{\rm scal}$
			& $\glc{}$ & $a_{\rm scal}$ & $b_{\rm scal}$
			&$\glc{}$ & $a_{\rm scal}$ & $b_{\rm scal}$\\
			\hline
			$\glc{4}(1234)$   & $4$ & $\frac{13}{3} $ 
			& $\glc{2}(13;24)$  & $\frac{4}{9}$ & $-\frac{2}{9}$ 
			& $\glc{22}(12,34)$ &  $-4$ & $ \frac{2}{3}$ \\
			\hline 
			$\glc{4}(1243)$   & $4$ & $\frac{13}{3} $ 
			& $\glc{2}(14;23)$  & $\frac{4}{9}$ & $-\frac{2}{9}$
			& $\glc{22}(13,24)$ & $\frac{2}{9} $ & $\frac{1}{18}$ \\
			\hline
			$\glc{4}(1324)$   & $-\frac{20}{9} $ &  $-\frac{11}{9} $ 
			& $\glc{2}(24;13)$  & $\frac{4}{9}$ & $-\frac{2}{9}$ 
			& $\glc{22}(14,23)$ & $\frac{2}{9} $ & $\frac{1}{18}$ \\
			\hline
			$\glc{2}(12;34)$  & $-4$  & $-\frac{16}{3}$ 
			& $\glc{2}(23;14)$  & $\frac{4}{9}$ & $-\frac{2}{9}$ 
			& & &\\
			\hline
		\end{tabular}
		\caption{Contributions of individual terms in the worldline integrand to the double and single poles of the two-loop Maxwell term in Scalar QED. 
		The $a_{\rm scal}$ sum to zero, the $b_{\rm scal}$ to $2$, see (\ref{beta-reg}).}
		\label{table:ta2}
	\end{table}

Although all terms here are already individually gauge invariant, the vanishing of the double pole still
happens through an intricate cancellation between them. Similarly, apart from the vanishing of the three-cycle contributions 
the other terms all contribute to the single pole, without any clear pattern emerging as one might have hoped.

\subsection{Spinor QED $\beta$-function}
\label{spinor-beta}

For the more relevant spinor QED case, let us also write down the individual contributions in terms of the $Y_{nl}$:
	\begin{equation}
		\begin{split}
			 \hat{\Gamma}^{(2)4}_{\rm spin}(1234)  = \hat{\Gamma}^{(2)4}_{\rm spin}(3124) &= - \frac{8(D-1)}{3D} \left(Y_{41} + 2Y_{42}\right){\rm tr}(f_3f_4)\,,\\
			 \hat{\Gamma}^{(2)4}_{\rm spin}(2314) & = - \frac{8}{9D} \left(2Y_{40} - 6Y_{41} - 9Y_{42}\right){\rm tr}(f_3f_4)\,,\\
			 \hat{\Gamma}^{(2)2}_{\rm spin}(12;34) & =  \frac{8}{3D} (D-1)\, k^2\, Y_{53}\,{\rm tr}(f_3f_4)\,,\\
			 \hat{\Gamma}^{(2)2}_{\rm spin}(13;24)  = \hat{\Gamma}^{(2)2}_{\rm spin}(23;14) & = \frac{2}{9D} \left(Y_{40} - 6Y_{41}\right){\rm tr}(f_3f_4)\,,\\
			 \hat{\Gamma}^{(2)2}_{\rm spin}(14;23)  = \hat{\Gamma}^{(2)2}_{\rm spin}(24;13) & = \frac{2}{9D} \left(Y_{40} - 6Y_{41}\right){\rm tr}(f_3f_4)\,,\\
			 \hat{\Gamma}^{(2)22}_{\rm spin}(12,34) & = \frac{4}{3}(D-1)\,Y_{41} \, {\rm tr}(f_3f_4)\,,\\
			 \hat{\Gamma}^{(2)22}_{\rm spin}(13,24)  = \hat{\Gamma}^{(2)22}_{\rm spin}(14,23) & = \frac{4}{9D} \, Y_{40}\,{\rm tr}(f_3f_4)\,.
		\end{split}
		\label{gamma-contri}
	\end{equation}
	Collecting all terms, we obtain
	\begin{equation}
		\begin{split}
			\Gamma^{(2)}_{\rm spin}(f_3,f_4)&=-\frac{e^4}{(4\pi)^{\frac{D}{2}}}\int\frac{d^Dk}{(2\pi)^D}\bigg[\frac{4}{3D}(D-4)(D-1)Y_{41} -\frac{8}{3D}(4D-7)Y_{42}\\
			&\hspace{3cm}+\frac{8}{3D}(D-1)\,k^2\, Y_{53}\bigg]{\rm tr}
			(f_3f_4)\,.
		\end{split}
		\label{gammas-beta}
	\end{equation}
As in the scalar case, we define
 	\begin{equation}
 		\Gamma^{(2)}_{\rm spin}(f_3,f_4)\equiv -\frac{e^4}{(4\pi)^{4}}\, \gamma_{\sq} (D,m^2) \,{\rm tr} (f_3f_4)\,.
 		\label{spin-beta1}
 	\end{equation}
Application of \eqref{u-integrals} leads to the following representation for $\gamma_{\rm spin} (D,m^2)$,
	\begin{equation}
		\begin{split}
			\gamma_{\rm spin} (D,m^2) &= \frac{(4 \pi)^{4}  \Gamma (4-D)}{4(4 \pi)^{D}m^{2(4-D)}} \, \int_0^1 \,du \, \frac{16}{3D}
			\Big\{ (D -4) (D -1) \Big[G(u,0)\Big]^{1-\frac{D}{2}} \\
			& \quad + (D -7) (D -2) \Big[G(u,0)\Big]^{2-\frac{D}{2}}\Big\}  .
		\end{split}
		\label{2loopuint}
	\end{equation}
This result agrees already at the integrand level with Eq. (36) of \cite{Schmidt:1994aq}, providing another excellent check on many of the equations of the present paper. 
Performing the $u$ - integral brings us to our final result for the induced Maxwell term,
  \begin{equation}
  	\begin{split}
  		\gamma_{\sq} (D,m^2)&= \frac{(4\pi)^{4}\Gamma\left(4-D\right)}{(4\pi)^D \, m^{2(4-D)}} \,  B\left(2-\frac{D}{2},2-\frac{D}{2}\right) \frac{(D-4)[34+D(5D-33)]}{3D(D-5)}\,.
  	\end{split}
  \end{equation}
  Note once more the global factor of $(D-4)$, which reduces the pole in dimensional regularization to a single one,
  \begin{equation}\label{coef-beta}
		\begin{split}
			\gamma_{\sq} (\epsilon,m^2)= \frac{6}{\epsilon} +\mathcal{O}(\epsilon^0)\,.
		\end{split}
	\end{equation}
	%
  
	%


Adding the appropriate term from mass renormalization \cite{Schmidt:1994aq}		%
	\bear
	\Delta\Gamma^{(2)}_{\rm spin}(f_3,f_4)\sim\frac{8e^4}{(4\pi)^4 \epsilon}{\rm tr}(f_3f_4) \,,
	\ear
we get the total pole of the induced Maxwell term,
	\bear
	\Gamma^{(2)}_{\rm spin}(f_3,f_4)+\Delta\Gamma^{(2)}_{\rm spin}(f_3,f_4)\sim \frac{2e^4}{(4\pi)^4\epsilon}{\rm tr}(f_3f_4)\,,\hspace{0.9cm}
	\ear
	which finally gives the correct two-loop spinor QED $\beta$-function \cite{joslut,Itzykson1980rh},
	\bear
	\beta^{(2)}_{\rm spin}(\alpha)=\frac{\alpha^3}{2\pi^2}\,.
	\ear
	Finally, in Table \ref{table:ta1} we list the double and single pole contributions of the individual terms in the worldline integrand. 
	The picture that emerges is similar to the scalar QED case above.

	\begin{table}[h!]
		\centering
		\begin{tabular}{ |c|c|c||c|c|c||c|c|c| }
			\hline
			$\gls{}$ & $a_{\rm spin}$ & $b_{\rm spin}$ 
			&$\gls{}$ & $a_{\rm spin}$ & $b_{\rm spin}$ 
			&$\gls{}$ & $a_{\rm spin}$ & $b_{\rm spin}$ \\
			\hline
			$\gls{4}(1234)$   & $-8$ & $\frac{10}{3} $
			& $\gls{2}(13;24)$  & $-\frac{8}{9}$ & $\frac{4}{9}$
			& $\gls{22}(12,34)$ &  $16$ & $ \frac{16}{3}$ \\
			\hline
			$\gls{4}(1243)$   & $-8$ & $\frac{10}{3} $
			& $\gls{2}(14;23)$  & $-\frac{8}{9}$ & $\frac{4}{9}$
			& $\gls{22}(13,24)$ & $\frac{8}{9} $ & $\frac{2}{9}$ \\
			\hline
			$\gls{4}(1324)$   & $\frac{16}{9} $ &  $-\frac{38}{9} $ 
			& $\gls{2}(24;13)$  & $-\frac{8}{9}$ & $\frac{4}{9}$ 
			& $\gls{22}(14,23)$ & $\frac{8}{9} $ & $\frac{2}{9}$ \\
			\hline 
			$\gls{2}(12;34)$  & $0$  & $-4$
			& $\gls{2}(23;14)$  & $-\frac{8}{9}$ & $\frac{4}{9}$
			& & &\\
			\hline
		\end{tabular}
		\caption{Contributions of individual terms in the worldline integrand to the double and single poles of the two-loop Maxwell term in Spinor QED. 
		The $a_{\rm spin}$ sum to zero, the $b_{\rm spin}$ to $6$, see (\ref{coef-beta}).}
		\label{table:ta1}
	\end{table}


	\subsection{Four-dimensional computation of the Spinor QED $\beta$-function}
	
	Although the calculation of corresponding quantities in Scalar and Spinor QED in the worldline formalism usually proceeds with only minor differences, 
	a surprising finding of \cite{Schmidt:1994aq} had been that, in the case of the two-loop QED $\beta$ functions, this holds true when employing dimensional regularization
	but not for strictly four-dimensional regularization schemes such as the use of a proper-time cutoff. In \cite{Schubert2001-73}, this observation was explained as a consequence
	of the above-mentioned theorem by Johnson et al. \cite{Johnson:1967pk} that the total induced effective action can have only a global divergence. 
	Since, unlike Scalar QED, Spinor QED does not possess true quadratic divergences, the required cancellation of subdivergences here leads to two constraint equations.
	And since, by the structure of the worldline representation of the induced Maxwell term, the integrand before the final integration over $u$ in $D=4$ must be of the form (compare \eqref{2loopuint})
		\begin{equation}
		\left(\frac{A}{G_{12}^2}+\frac{B}{G_{12}}+C\right){\rm tr}(F^2)\,,
	\end{equation}
one can first conclude from the absence of a quadratic subdivergence that $A = 0$, and then from the absence of a logarithmic subdivergence that $B = 0$,
leaving only the coefficient $C$ non-vanishing and making the final $u$-integral trivial (this argument does not work in dimensional regularization due to the principal
suppression of quadratic subdivergences by that scheme). 

Thus one would expect that also the present recalculation should simplify in the Spinor QED case when a four-dimensional scheme is used. 
And indeed, if we return to  \eqref{gammas-beta} and put $D=4$ there, we find 
\bear
	\gamma_{\sq} (D,m^2)&\xrightarrow[]{D=4}&(4\pi)^{2} \int\frac{d^4k}{(2\pi)^4}\bigg[- 6 Y_{42}+2Y_{53}k^2\bigg]\,,
\ear
and then using any four-dimensional method for the regularization of the global divergence that is still contained in the global $T$-integration will lead to a trivial $u$-integration.
%
%

\section{Delbr\"uck scattering at low energies}
\label{delbruecksection}
In this section, we compute the differential cross section of Delbr\"uck scattering in scalar and spinor QED under the assumption
that the photon that interacts with the Coulomb field has low energy. For the spinor QED case, 
this quantity was computed in detail in \cite{Costantini:1971cj}, therefore we will follow their conventions for easy comparison. 

\begin{figure}[h]
	\centering 
	\includegraphics[width=.4\textwidth]{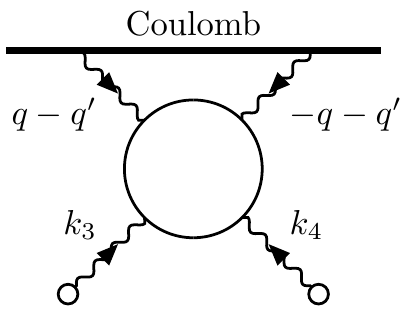}
	\caption{Feynman diagram for low-energy Delbr\"uck scattering.}
\end{figure}
Here we use our above results for the four-photon amplitudes with two low-energy photons, and replace the two unrestricted legs 
$1$ and $2$ with two Coulomb photons. 
Furthermore, we now take the two low-energy photons ($3$ and $4$) on-shell. 

The vector potential for the photons from the Coulomb field is given by
\begin{equation}
	A_\mu(x)= \left(-\frac{Ze}{4\pi r},\,{\bf 0}\right)\,,
\end{equation}
and has the following Fourier representation,
\begin{equation}
	A_\mu(x)=\varepsilon_\mu \int\frac{d^4k}{(2\pi)^{4}}\, \frac{Ze}{k^2}\, 2\pi\, \delta(k_0)\,\e^{ik\cdot x}\,.
\end{equation}
From here the new vertex operator for photons from the Coulomb field read as 
\begin{equation}
	\begin{split}
		V^\gamma_{\rm Nuc}[k,\varepsilon]&=\frac{Ze}{(2\pi)^3}\int d^4k\frac{\delta(k_{0})}{k^2}\int_0^Td\tau\, \varepsilon\cdot\dot{x}\e^{ik\cdot x}\\
		&=\frac{Ze}{(2\pi)^3}\int \frac{d^{3} {\bf k}}{{\bf k}^2}\int_0^Td\tau\, \varepsilon\cdot\dot{x}\,\e^{i{\bf k}\cdot {\bf x}}\,.
	\end{split}
\end{equation}
Thus in the present formalism, we have
\begin{equation}\label{DSscal}
	\Gamma_{\left\{\s  (34)\atop\sq  (34)\right\} } 
	= \left\{1\atop -2\right\} 
	\frac{1}{2}
	\frac{e^4 (Ze)^2}{(4\pi)^2(2\pi)^{6}} \int\frac{ (d^{3}{{\bf k}_1}) (d^{3}{{\bf k}_2})}{{\bf k}_1^2 {\bf k}_2^2}  (2\pi)^4 \delta^4(k_1 + k_2 + k_3 + k_4)  \hat{\Gamma}_{\left\{\s(34)\atop\sq(34)\right\}}\,,
\end{equation}
with $ \hat{\Gamma}_{\left\{\s(34)\atop\sq(34)\right\}}$ as defined in \eqref{decomp}, \eqref{Intu1u2T}.
Note the factor of $\frac{1}{2}$ that takes the symmetry between legs $1$ and $2$ into account
\footnote{In Feynman diagram calculations of Delbr\"uck scattering, this factor if usually implemented by restricting the sum over diagrams
to run only over three diagrams, instead of the six light-by-light scattering ones (which are equal in pairs by CP invariance). In the worldine formalism
we do not have this option.}
.
In order to reproduce the result given in \cite{Costantini:1971cj} for the differential cross section, we employ the following kinematics
\begin{equation}\label{k1k2k3k4}
\begin{split}
k_1&=(0,{\bf q}-{\bf q}')~,~~~~~~k_2=(0,-{\bf q}-{\bf q}'), \\
k_3&=(i\omega,{\bf k}+{\bf q}')~,~~~~~k_4=(-i\omega,{\bf q}'-{\bf k}),
\end{split}
\end{equation}
where
\begin{equation}
\begin{split}
{\bf q'}&=(\omega\sin\theta/2,0,0),\\
{\bf k}&=(0,0,\omega\cos\theta/2),\\
{\bf q}&=(q_1,q_2,q_3),
\end{split}
\end{equation}
with $\omega$ as the energy and $\theta$ the scattering angle. The polarizations are chosen as
\begin{equation}
\begin{split}
\varepsilon_1^\mu &= \varepsilon_2^\mu = (i,0,0,0),\\
\varepsilon_3^\mu &= \frac{1}{\sqrt{2}}(0,-i\lambda_3\cos\theta/2,1,+i\lambda_3\sin\theta/2),\\
\varepsilon_4^\mu &= \frac{1}{\sqrt{2}}(0,-i\lambda_4\cos\theta/2,1,-i\lambda_4\sin\theta/2),
\end{split}
\end{equation}
where $\lambda_i=\pm 1$ for right-and left-handed circular polarization, respectively.
In the following it is understood that $\hat{\Gamma}_{\s(34)}^{+-} = \hat{\Gamma}_{\s(34)}|_{\lambda_3 = 1,\, \lambda_4 =-1}$ etc. and we will also use the abbreviations
\begin{equation}
	P_0 \equiv \frac{{\rm arcsinh}\left(\frac{q}{2m}\right) }{q\sqrt{4m^2 + q^2}}, \qquad S \equiv \sin\frac{\theta}{2}, \qquad C \equiv \cos\frac{\theta}{2}\,.
\end{equation}
With the kinematics of (\ref{k1k2k3k4}) and using conservation of momentum we can write (\ref{DSscal}) as 
\begin{equation}
\Gamma_{\left\{\s  (34)\atop\sq  (34)\right\}}
=   \left\{1\atop -2\right\}
\frac{1}{2}
 \frac{e^4 (Ze)^2}{(4\pi)^2(2\pi)^{6}} (2\pi)^4 \delta(k_3^0 + k_4^0) \int\frac{ d^{3}{\bf q}}{\vert{\bf q}-{\bf q'}\vert^2\vert{\bf q}+{\bf q}'\vert^2}\hat{\Gamma}_{\left\{\s(34)\atop\sq(34)\right\}} \,.
\end{equation}
For convenience, let us further define
\begin{equation}
	\tilde{\Gamma}_{\left\{\s  \atop\sq  \right\}}
	\equiv \int\frac{ d^{3}{\bf q}}{\vert{\bf q}-{\bf q'}\vert^2\vert{\bf q}+{\bf q}'\vert^2} \hat{\Gamma}_{\left\{\s(34)\atop\sq(34)\right\}}\,.
\end{equation}
Since we are considering the low-energy case, $\omega \ll m$, we neglect contributions of order superior to $\omega^2$. We notice that
\begin{equation}
	\begin{split}
		\int \frac{ d^{3}{\bf q}}{\vert{\bf q}-{\bf q'}\vert^2\vert{\bf q}+{\bf q}'\vert^2} & = \int_0^\infty \int_0^\pi \int_0^{2\pi} \frac{q^2 \sin \theta' dq \, d\theta' \, d\phi'}{\left( q^2 + \omega^2 \sin^2 \frac{\theta}{2}\right)^2 - 4q_1^2 \omega^2 \sin^2 \frac{\theta}{2}}\\
		& = \int_0^\infty \int_0^\pi \int_0^{2\pi} \frac{\sin \theta' dq \, d\theta' \, d\phi'}{q^2} + \mathcal{O}(\omega)\,.
	\end{split}
\end{equation}
Using the kinematics of (\ref{k1k2k3k4}), we find 
\begin{equation}
	\begin{split}
		\hat{\Gamma}_{\s(34)}^{+-} &= \frac{4 \omega^2}{3m^2 q^4 (4m^2 + q^2)}\Big\{3m^2[(6m^2 + q^2) - 8m^2(3m^2 + q^2)P_0](q_2^2 - q_1^2)\\
		&\quad + [q^4 (2m^2 + q^2) -3m^2(6m^2+q^2)q_2^2] S^2  - 8m^4[ q^4 - 3(3m^2+q^2)q_2^2]S^2P_0\Big\} + \mathcal{O}(\omega^3)\,,
	\end{split}
\end{equation}
\begin{equation}
	\begin{split}
		\hat{\Gamma}_{\sq(34)}^{+-} &= \frac{4 \omega^2}{3 q^4 (4m^2 + q^2)^2}\Big\{3(4m^2 + q^2)[(6m^2+q^2)- 8m^2 (3 m^2 + q^2)P_0](q_2^2 - q_1^2) \\
		&\quad + [ q^4(2m^2 - q^2) -3(4m^2+q^2)(6m^2+q^2)q_2^2]S^2 \\
		&\quad -8m^2 [ q^4(m^2 + q^2)- 3(4 m^2 + q^2) (3 m^2 + q^2)q_2^2]S^2 P_0\Big\} + \mathcal{O}(\omega^3)\,,
	\end{split}
\end{equation}
for the helicity non-conserving component, and
	\begin{equation}
		\begin{split}
			\hat{\Gamma}_{\s(34)}^{++} &= \frac{4 \omega^2}{3 q^4 (4m^2 + q^2)}\Big\{4[(6m^2 + q^2)- 8m^2 (3m^2 +q^2)P_0](q_2^2 - q_3^2) \\
			&\quad- [3q^2(2m^2 + q^2) + 4(6m^2+q^2)q_2^2] C^2 \\
			&\quad +4[q^2(6m^4 + 4m^2 q^2 + q^4)+ 8m^2(3m^2+ q^2)q_2^2]C^2P_0\Big\}  + \mathcal{O}(\omega^3)\,,
		\end{split}
	\end{equation}
	\begin{equation}
		\begin{split}
			\hat{\Gamma}_{\sq(34)}^{++} &= \frac{4 \omega^2}{3 q^4 (4m^2 + q^2)^2}\Big\{(6 m^2 + q^2) (16 m^2 + 7 q^2)(q_2^2 - q_3^2) \\
			&\quad+8[(3 m^2 +  q^2) (4 m^2 + q^2)^2  - 3  m^4 q^2](q_3^2 - q_2^2)P_0 \\
			&\quad- [3q^2(8m^4  -2m^2q^2 - q^4) + (6 m^2 + q^2) (16 m^2 + 7 q^2)q_2^2]C^2 \\
			&\quad+ 8q^2 (12 m^6  - m^4 q^2 - 5 m^2 q^4 - q^6 ) C^2 P_0\\
			&\quad+ 8 [(3 m^2 +  q^2) (4 m^2 + q^2)^2  - 3  m^4 q^2]q_2^2 C^2 P_0\Big\}  + \mathcal{O}(\omega^3)\,,
		\end{split}
	\end{equation}
for the conserving one.  
To perform the integral over ${\bf q}$ we use spherical coordinates:
\begin{equation}
	q_1 = q \cos \theta', \qquad q_2 = q \sin \theta' \cos \phi', \qquad q_3 = q \sin \theta' \sin \phi'\,.
\end{equation}
The integrals over $\theta'$ and $\phi'$ are trivial, and what remains to be calculated is
\begin{equation}
	\tilde{\Gamma}_{\s}^{+-} = \frac{16 \pi S^2 \omega^2 }{3}
	 \int_0^\infty \frac{ dq}{q^2} \, \frac{-6m^4 + m^2 q^2 + q^4 + 24 m^6 P_0}{ m^2 q^2 (4m^2 + q^2)}\,,
\end{equation}
\begin{equation}
	\tilde{\Gamma}_{\sq}^{ +-}  = \frac{32 \pi S^2 \omega^2 }{3}
	 \int_0^\infty \frac{ dq}{q^2} \, \frac{-12 m^4 - 4 m^2 q^2 - q^4 + 24 m^4(2m^2 + q^2) P_0}{ q^2 (4m^2 + q^2)^2}\,,
\end{equation}
\begin{equation}
	\tilde{\Gamma}_{\s}^{ ++}  = \frac{16 \pi C^2 \omega^2 }{9}
	 \int_0^\infty \frac{ dq}{q^2} \, \frac{-42m^2 -13 q^2 + 4(42 m^4 + 20 m^2 q^2 + 3 q^4) P_0}{q^2 (4m^2 + q^2)}\,,
\end{equation}
\begin{equation}
	\tilde{\Gamma}_{\sq}^{++}  = \frac{32 \pi C^2 \omega^2 }{9}
	 \int_0^\infty \frac{ dq}{q^2} \, \frac{-84 m^4 -20 m^2 q^2 +q^4 + 8(42 m^6 + 17 m^4 q^2 -2m^2 q^4 - q^6) P_0}{q^2 (4m^2 + q^2)^2}\,.
\end{equation}
Performing the integral over $q$, we get
\begin{equation}
	\tilde{\Gamma}_{\s}^{ +-} = \frac{15 \pi^3 S^2 \omega^2 }{32 m^3}, \qquad
	\tilde{\Gamma}_{\sq}^{ +-}  = -\frac{5 \pi^3 S^2 \omega^2 }{32 m^3}\,,
\end{equation}
\begin{equation}
	\tilde{\Gamma}_{\s}^{ ++} = \frac{3 \pi^3 C^2 \omega^2 }{32 m^3}, \qquad
	\tilde{\Gamma}_{\sq}^{ ++}  = -\frac{73 \pi^3 C^2 \omega^2 }{288 m^3}\,.
\end{equation}

Finally, the differential cross section is
\begin{equation}
	d\sigma_{\s (\lambda_3 \lambda_4)}  = \frac{(Z\alpha)^4 \alpha^2}{4(2\pi)^{6}} \, \vert \tilde{\Gamma}_{\s }^{\lambda_3 \lambda_4} \vert^2 \, d\Omega\,,
\end{equation}
\begin{equation}
	d\sigma_{\sq (\lambda_3 \lambda_4)}  =  \frac{(Z\alpha)^4 \alpha^2}{(2\pi)^{6}} \, \vert \tilde{\Gamma}_{\sq}^{\lambda_3 \lambda_4} \vert^2 \, d\Omega\,.
\end{equation}

For scalar QED, we have
\begin{eqnarray}
	d\sigma_{\s(++)} &=& d\sigma_{\s(--)} = (Z\alpha)^4 \left(\frac{\alpha}{m}\right)^2 \left(\frac{3}{16}\right)^2 \left(\frac{1}{32}\right)^2 \left(\frac{\omega}{m}\right)^4 \cos^4 \frac{\theta}{2} d\Omega\,,\\
	d\sigma_{\s(+-)} &=& d\sigma_{\s(-+)} = (Z\alpha)^4 \left(\frac{\alpha}{m}\right)^2 \left(\frac{15}{16}\right)^2 \left(\frac{1}{32}\right)^2 \left(\frac{\omega}{m}\right)^4 \sin^4 \frac{\theta}{2} d\Omega\,.
\end{eqnarray}

For spinor QED, we find
\begin{eqnarray}
	d\sigma_{\sq(++)} &=& d\sigma_{\sq(--)} = (Z\alpha)^4 \left(\frac{\alpha}{m}\right)^2 \left(\frac{73}{72}\right)^2 \left(\frac{1}{32}\right)^2 \left(\frac{\omega}{m}\right)^4 \cos^4 \frac{\theta}{2} d\Omega\,,\\
	d\sigma_{\sq(+-)} &=& d\sigma_{\sq(-+)} = (Z\alpha)^4 \left(\frac{\alpha}{m}\right)^2 \left(\frac{5}{8}\right)^2 \left(\frac{1}{32}\right)^2 \left(\frac{\omega}{m}\right)^4 \sin^4 \frac{\theta}{2} d\Omega\,,
\end{eqnarray}
in agreement with \cite{Costantini:1971cj}.

\section{Summary and outlook}
\label{summary}

In this second part of our series of papers on the off-shell four-photon amplitudes in scalar and spinor QED
we have obtained these amplitudes  for the case where two of the photon momenta were taken in the low-energy limit,
as defined in part I. We have used the worldline representation of these amplitudes which, as outlined in part I, allows
one to treat the scalar and spinor cases in parallel, and to arrive at a permutation and gauge-invariant decomposition of these amplitudes 
that is closely related to the one previously obtained by \cite{Costantini:1971cj}. 
The coefficient functions in this decomposition are written in terms of Feynman-Schwinger type parameter integrals, and as our main result
we have explicitly evaluated them both in four and in $D$ dimensions. Since in our formalism the four-photon amplitudes are manifestly free
of UV divergences, the former results are sufficient for one-loop applications. The latter formulas serve the double purpose of making the off-shell
four-photon amplitudes useful as input for higher-loop calculations in dimensional regularization, but also giving the correct results for them in other 
physical dimensions (arbitrary $D$ for scalar QED,  even $D$ for spinor QED). 
After the integrating out of the two low-energy legs these integrals are of two-point
type, and can therefore be written in terms of ${}_2F_1$ for general $D$, and  for $D=4$ in terms of trigonometric functions.

To the best of our knowledge, for this momentum configuration the four-photon amplitudes have not been obtained before. 
However, for the special case $k_1+k_2=k_3+k_4= 0$ they can be extracted from the low-energy limit of the known
vacuum polarization tensors in a constant field, and performing this comparison provided an excellent check on our results for both scalar and spinor QED.

From a technical point of view, probably the most important aspect of our calculation is that it demonstrates how to integrate out a low-energy photon leg 
{\sl without fixing an ordering for the remaining legs}. As discussed in part I, when using the light-by-light diagram
as a subdiagram in higher-loop calculation this property effectively allows one to unify the calculation of Feynman diagrams of different topologies,
and considering the proliferation of diagrams in higher-loop QED calculations it is obviously of great interest to explore what level of simplification can
be achieved along these lines. 

Off-shell photon legs can be used for creating internal photons by sewing, or for connecting to external fields. 
As an example for sewing, we have used our results for a construction of the two-loop vacuum polarisation tensors in the low-energy limit, which allowed
us to recover the two-loop $\beta$-function coefficients for scalar and spinor QED. Although by present-days standards this is easy enough to do using Feynman diagrams,
the worldline calculation in the version presented here has the advantage that it replaces the non-gauge invariant decomposition into diagrams by a gauge-invariant
decomposition into sixteen substructures. This leads not only to a more compact parameter integral representation, but also has given us an opportunity to probe into
the nature of the extensive cancellations that one generally finds in perturbative QED calculations \cite{Johnson:1967pk,Cvitanovic:1977dp,Broadhurst:1995dq,Huet:2017ydx}. 
If, as is sometimes assumed, these were entirely due to 
gauge invariance, one might have expected the cancellation of the double poles in the scalar and spinor QED $\beta$-function calculations to have happened 
already at the level of each of the gauge-invariant partial structures; instead, we have seen an intricate pattern of cancellations between the various structures. 
Another interesting result of this calculation has been that four of the sixteen partial structures, namely the one involving a three-cycle, drop out in the sewing.
It remains to be seen whether this fact admits a generalization to larger numbers of external photons and multiple sewing. 

As an example of an external-field calculation, we have applied our formulas to a calculation of the low-energy Delbr\"uck scattering cross sections for both scalar and spinor QED.  
For the spinor QED case this served just as another check of efficiency and correctness, while the scalar QED result is new, to the best of our knowledge.
Both the scalar and spinor QED calculations have been presented in a way that would be easy to adapt to other external fields. 

The forthcoming third part of this series is devoted to the explicit, $D$-dimensional calculation of the off-shell four-photon amplitudes with only 
one leg taken in the low-energy limit and its applications. 

\paragraph{Acknowledgments:} We would like to thank D. Broadhurst, F. Karbstein and D. Kreimer for discussions and correspondence. 
C. Lopez-Arcos and M. A. Lopez-Lopez thank CONACYT for financial support. N. Ahmadiniaz and M. A. Lopez-Lopez would like to thank R. Sch\"utzhold for his support. 


\appendix

\section{Collection of integral formulas}\label{app2}

Here we collect a number of results for the parameter integrals appearing in one-loop worldline calculations, 
mostly taken from \cite{Schubert2001-73}. All Green's functions have been rescaled to the unit circle, $T=1$. 

\subsection{Integrating out a low-energy leg}

All integrals encountered in the integrating out of a low-energy photon can, using the identities \eqref{idbasic},
be reduced to integrals over polynomials in $\dot G_{ij}$. For this type of integrals a
closed-form master formula is available even for the most general (abelian) case, 
integrating an arbitrary monomial in the $\dot G_{ij}$'s involving an arbitrary number of variables
in one of the variables, and giving the result as a polynomial in the remaining $\dot G_{ij}$'s 
\cite{Edwards:2021elz}:
\bear
\int_0^1 du\,
\dot G(u,u_1)^{k_1}
\dot G(u,u_2)^{k_2}
\cdots
\dot G(u,u_n)^{k_n}
&=&
{1\over 2n}
\sum_{i=1}^n\,
\prod_{j\ne i}
\sum_{l_j=0}^{k_j}
{k_j\choose l_j}
\dot G_{ij}^{k_j-l_j}
\sum_{l_i=0}^{k_i}{k_i\choose l_i}
\nonumber\\&&\hspace{-180pt}\times
{(-1)^{\sum_{j=1}^n l_j}
\over (1+\sum_{j=1}^n l_j)n^{\sum_{j=1}^n l_j}}
\biggr\lbrace
\Bigl( \sum_{j\ne i}\dot G_{ij} +1 \Bigr)
^{1+\sum_{j=1}^n l_j}
- (-1)^{k_i-l_i}
\Bigl(
\sum_{j\ne i}\dot G_{ij} -1
\Bigr)^{1+\sum_{j=1}^n l_j }
\biggr\rbrace
\, .
\nonumber\\
\label{intmaster}
\ear
At the four-photon level, apart from the single four-point integral
\begin{equation}\label{int6}
\int_0^1 du_4\,  \dot{G}_{41} \dot{G}_{42} \dot{G}_{43} = - \frac{1}{6}(\dot{G}_{12}-\dot{G}_{23})(\dot{G}_{23}-\dot{G}_{31})(\dot{G}_{31}-\dot{G}_{12})\,,
\end{equation}
only three-point integrals appear, for which the master formula \eqref{intmaster} reduces to 
\begin{equation}\label{int7}
\int_0^1 du_3 \,\dot{G}_{13}^k \dot{G}_{32}^l= \frac{k!l!}{2}\sum_{m=0}^l \frac{(1-(-1)^{k+l-m+1})\dot{G}_{12}^m - (1-(-1)^m)\dot{G}_{12}^{k+l-m+1}  }{m!(k+l-m+1)!}\,.
\end{equation}
However, the latter formula generally does not provide the most compact way of writing the result in terms of the $\dot G_{ij}$, therefore for easy reference we
provide below a list of optimized versions for all the integrals that appeared in our calculations.  
\begin{equation}
	\begin{split}\label{intGdots}
		&\int_0^1 du_3 \, \dot{G}_{13} = 0\,, \\
		&\int_0^1 du_3 \, \dot{G}_{13}^2 = \frac{1}{3}\,, \\
		&\int_0^1 du_3 \, \dot{G}_{13}^4 = \frac{1}{5}\,, \\
		&\int_0^1 du_3 \dot{G}_{13} \dot{G}_{32} = \frac{1}{6} - \frac{1}{2}\dot{G}_{12}^2 \,,  \\
		&\int_0^1 du_3 \, \dot{G}_{13} \dot{G}_{32}^2 =\frac{1}{3}(\dot{G}_{12} - \dot{G}_{12}^{3})\,, \\
		&\int_0^1 du_3 \, \dot{G}_{23}^3 \dot{G}_{31}  = \frac{3!}{5!}-\frac{\dot{G}_{12}^{4}}{4}\,, \\
		&\int_0^1 du_3 \, \dot{G}_{13}^2 \dot{G}_{32}^2 =\frac{4}{5!}-\frac{\dot{G}_{12}^{4}}{3!} + \frac{\dot{G}_{12}^{2}}{3!} \,, \\
		&\int_0^1 du_3 \, \dot{G}_{13}^4 \dot{G}_{32}  = \frac{1}{5}(\dot{G}_{12} - \dot{G}_{12}^{5}) \,,\\
		&\int_0^1 du_3 \, \dot{G}_{13}^3 \dot{G}_{32}^2  = \frac{1}{10}(\dot{G}_{12} - \dot{G}_{12}^{5})\,,
	\end{split}
\end{equation}
\begin{equation}\label{intGs}
	\begin{split}
		&\int_0^1 du_3 \, G_{13} = \frac{1}{6}\,, \\
		&\int_0^1 du_3 \, G_{13} G_{32} = \frac{1}{30} - \frac{1}{6} G_{12}^2  \,, \\
		&\int_0^1 du_3 \, G_{23} \dot{G}_{31}  = \frac{1}{3}\dot{G}_{12}G_{12}\,.
	\end{split}
\end{equation}

\subsection{The functions $Y_{nl}$ and tensor reduction}\label{B.2}

Here we study some properties of the functions $Y_{nl}$ defined in \eqref{defYnl},
\begin{equation}\label{def2Ynl}
Y_{nl} = \int_0^{\infty} \frac{dT}{T}T^{n-D/2} ~\int_0^1 du_1\int_0^1 du_2 ~ G_{12}^l \e^{-T[m^2 - G_{12} k_1 \cdot k_2]} \,.
\end{equation}
In terms of  $\hat{k}_{12} = \frac{k_1 \cdot k_2}{m^2} $ and using the translational invariance to fix $u_2=0$ and  $u_1=u$, so that
$G_{12} = u(1-u)$, we can perform the following tensor reduction 
\bear
Y_{nl} &=& \int_0^{\infty} \frac{dT}{T}T^{n-D/2} ~\int_0^1 du ~ [u(1-u)]^l \e^{-Tm^2[1 - u(1-u)\hat{k}_{12}]}  \nonumber\\
 &=& \frac{\Gamma\left(n-\frac{D}{2}\right)}{m^{2n-D}} ~\int_0^1 du ~ \frac{[u(1-u)]^l}{[1 - u(1-u) \hat{k}_{12} ]^{n-\frac{D}{2}}} \nonumber\\
&=& \frac{\Gamma\left(n-\frac{D}{2}-l\right)}{m^{2n-D} } ~\frac{d^l}{d\hat{k}_{12}^l}\int_0^1 du ~ \frac{1}{[1 - u(1-u)\hat{k}_{12}]^{n-l-\frac{D}{2}}} \, .
\label{Ynlcalc}
\ear
This leaves us with the single integral
\begin{equation}
\begin{split}
\int_0^1 du ~ \frac{1}{[1 - u(1-u)\hat{k}_{12}]^{n-l-\frac{D}{2}}} = {}_2F_1\left(1,n-l-\frac{D}{2};\frac{3}{2};\frac{\hat{k}_{12}}{4} \right) \, .
\end{split}
\end{equation}
So, in terms of the gaussian hypergeometric function ${}_2F_1$,
\begin{equation}\label{Ynl_2F1}
\begin{split}
Y_{nl} = \frac{\Gamma\left(n-\frac{D}{2}-l\right)}{m^{2n-D} } ~\frac{d^l}{d\hat{k}_{12}^l} ~ {}_2F_1 \left(1,n-l-\frac{D}{2}; \frac{3}{2};\frac{\hat{k}_{12}}{4} \right) \, .
\end{split}
\end{equation}

In Section \ref{section-beta} we need the elementary integral 
	\bear\label{u-integrals}
		\int\frac{d^Dk}{(2\pi)^D}\, (k^2)^\lambda \, Y_{nl}
		& = & \,\frac{\Gamma\left(\frac{D}{2}+\lambda \right) \, \Gamma\left(n-\lambda-D\right)}{(4\pi)^{\frac{D}{2}}\Gamma\left(\frac{D}{2}\right) m^{2(n-\lambda-D)}}
		\, \int_0^1 \,du \bigl[u(1-u)\bigr]^{l-\lambda-\frac{D}{2}}
		\nonumber\\
		& = & \,\frac{\Gamma\left(\frac{D}{2}+\lambda \right) \, \Gamma\left(n-\lambda-D\right)}{(4\pi)^{\frac{D}{2}}\Gamma\left(\frac{D}{2}\right) m^{2(n-\lambda-D)}}
		\, \, B \Bigl(l-\lambda-\frac{D}{2}+1,l-\lambda-\frac{D}{2}+1\Bigr).
		\nonumber\\
	\ear

	The following identities are not used in the present paper, but let us include them here for whatever their worth might be:
\begin{equation}
	Y_{nl} = \Gamma\left(n-\frac{D}{2}\right) \hat{k}_{12}^{-l}\, \sum_{j=0}^{l} (-1)^j \binom{l}{j} \left[\Gamma\left(n -j -\frac{D}{2}\right)\right]^{-1} m^{-2j} \, Y_{(n-j)0} \, ,
	\label{YnltoYn0}
\end{equation}
which can be used to recursively eliminate all $Y_{nl}$'s with $l\ne 0$. And
\begin{equation}
	(2n-D)(2n-D -1)\, Y_{n0} =  \left[2 -(2n-D) \left(\hat{k}_{12}^2-8\right)\right]\, m^2 \, Y_{(n+1)0} +  \left(\hat{k}_{12}^2-4\right) m^4\,  Y_{(n+2)0}\,,
\end{equation}
that follows from the following identity of contiguous functions \cite{Stegun}
\begin{equation}
	(c-b){}_2F_1(a,b-1;c;z) + (2b-c-bz+az) {}_2F_1(a,b;c;z) + b (z-1) {}_2F_1(a,b+1;c;z) = 0.
\end{equation}

\section{Explicit results for $\Qcc$ and $\Qss$}
\label{app4}
In this appendix, we explicitly write down the results for $Q_{\s}$ and $Q_{\sq}$ after integration over $u_4$ and $u_3$, under the assumption that photons $3$ and $4$ are taken in the low-energy limit.

\subsection{Scalar QED}\label{Q(34)scal}

The explicit form of $\Qcc$ is written as
\begin{equation}
\Qcc^4(1234) = \frac{2}{3} Z_4(1234) \left(G_{12}-4G_{12}^2\right) ,
\end{equation}
\begin{equation}
\Qcc^4(2314) = \frac{1}{9} Z_4(2314)\left(1-12 G_{12} + 36 G_{12}^2 \right),
\end{equation}
\begin{equation}
\Qcc^4(3124) = \frac{2}{3} Z_4(3124) \left(G_{12}-4G_{12}^2\right)  ,
\end{equation}
\begin{equation}
\Qcc^3(123;4) = -\frac{T}{9} Z_3(123) k_2 \cdot f_4 \cdot k_1 \left(G_{12} - 10 G_{12}^2 + 24 G_{12}^3\right),
\end{equation}
\begin{equation}
\Qcc^3(234;1) = 0,
\end{equation}
\begin{equation}
\Qcc^3(341;2) = 0,
\end{equation}
\begin{equation}
\Qcc^3(412;3) = -\frac{T}{9} Z_3(412) k_2 \cdot f_3 \cdot k_1 \left(G_{12} - 10 G_{12}^2 + 24 G_{12}^3\right),
\end{equation}
\begin{equation}
\begin{split}
\Qcc^2(12;34) &= -\frac{T}{18} Z_2(12)  \left(1- 4G_{12} \right)   \Bigg\{\Bigg[\frac{1}{5} k_1 \cdot f_3 \cdot f_4 \cdot k_1  + \left(\frac{1}{5} - 6 G_{12}^2 \right) k_1 \cdot f_3 \cdot f_4 \cdot k_2\\
&\quad + (1\leftrightarrow 2) \Bigg] + 2T \left(1- 4G_{12} \right)G_{12}^2 \, k_1 \cdot f_3 \cdot k_2 k_1 \cdot f_4 \cdot k_2  \Bigg\} \,,
\end{split}
\end{equation}
\begin{equation}
\Qcc^2(13;24) + \Qcc^2(23;14) = -\frac{T}{9} Z_2(13) k_1 \cdot f_2 \cdot f_4 \cdot k_1 \,  \left(1- 4G_{12} \right)G_{12} + (1\leftrightarrow 2) \,,
\end{equation}
\begin{equation}
\Qcc^2(14;23) + \Qcc^2(24;13)= - \frac{T}{9} Z_2(14) k_1 \cdot f_2 \cdot f_3 \cdot k_1 \,  \left(1- 4G_{12} \right)G_{12} + (1\leftrightarrow 2) \,,
\end{equation}
\begin{equation}
\Qcc^2(34;12) = 0,
\end{equation}
\begin{equation}
\Qcc^{22}(12,34) = \frac{1}{3}Z_2(12) Z_2(34) \left(1-4G_{12}\right) ,
\end{equation}
\begin{equation}
\Qcc^{22}(13,24) =\frac{1}{9} Z_2(13)Z_2(24),
\end{equation}
\begin{equation}
\Qcc^{22}(14,23) = \frac{1}{9} Z_2(14) Z_2(23).
\end{equation}

\subsection{Spinor QED}\label{Q(34)spin}

The explicit form of $\Qss$ is written as
\begin{equation}
\Qss^4(1234) = - \frac{4}{3} Z_4(1234) \left(G_{12}+2G_{12}^2 \right) ,
\end{equation}
\begin{equation}
\Qss^4(2314) = - \frac{4}{9} Z_4(2314) \left(2- 6G_{12} - 9 G_{12}^2 \right),
\end{equation}
\begin{equation}
\Qss^4(3124) = - \frac{4}{3} Z_4(3124) \left(G_{12}+2G_{12}^2 \right) ,
\end{equation}
\begin{equation}
\Qss^3(123;4) = \frac{2T}{9} Z_3(123) k_2 \cdot f_4 \cdot k_1 \left(G_{12} - G_{12}^2 - 12G_{12}^3\right),
\end{equation}
\begin{equation}
\Qss^3(234;1) = 0,
\end{equation}
\begin{equation}
\Qss^3(341;2) = 0,
\end{equation}
\begin{equation}
\Qss^3(412;3) = \frac{2T}{9} Z_3(412) k_2 \cdot f_3 \cdot k_1 \left(G_{12} - G_{12}^2 - 12G_{12}^3\right),
\end{equation}
\begin{equation}
\begin{split}
\Qss^2(12;34) &= \frac{2T}{9} Z_2(12) \,G_{12} \Bigg\{\Bigg[\frac{1}{5} k_1 \cdot f_3 \cdot f_4 \cdot k_1  + \left(\frac{1}{5} - 6 G_{12}^2  \right) k_1 \cdot f_3 \cdot f_4 \cdot k_2 + (1\leftrightarrow 2)\Bigg]\\
&\quad + 2T\, \left(1- 4G_{12} \right)G_{12}^2 \, k_1 \cdot f_3 \cdot k_2 k_1 \cdot f_4 \cdot k_2 \Bigg\} \,,
\end{split}
\end{equation}
\begin{equation}
\Qss^2(13;24) + \Qss^2(23;14) = \frac{2T}{9} Z_2(13) k_1 \cdot f_2 \cdot f_4 \cdot k_1 \,  \left(1- 4G_{12} \right)G_{12} + (1\leftrightarrow 2)\,,
\end{equation}
\begin{equation}
\Qss^2(14;23) + \Qss^2(24;13)= \frac{2T}{9} Z_2(14) k_1 \cdot f_2 \cdot f_3 \cdot k_1 \,  \left(1- 4G_{12} \right)G_{12} + (1\leftrightarrow 2) \,,
\end{equation} 
\begin{equation}
\Qss^2(34;12) = 0,
\end{equation} 
\begin{equation}
\Qss^{22}(12,34) = \frac{8}{3} Z_2(12) Z_2(34) G_{12}\, ,
\end{equation}
\begin{equation}
\Qss^{22}(13,24) = \frac{4}{9} Z_2(13)Z_2(24),
\end{equation}
\begin{equation}
\Qss^{22}(14,23) = \frac{4}{9} Z_2(14)Z_2(23) .
\end{equation}


\end{document}